\documentclass[Afour,sageh,times]{sagej}

\usepackage{moreverb,url}

\usepackage[colorlinks,bookmarksopen,bookmarksnumbered,citecolor=black,urlcolor=black]{hyperref}

\newcommand\BibTeX{{\rmfamily B\kern-.05em \textsc{i\kern-.025em b}\kern-.08em
T\kern-.1667em\lower.7ex\hbox{E}\kern-.125emX}}

\def\volumeyear{2024}

\usepackage{multirow}
\usepackage{makecell}
\usepackage{tabularx}
\usepackage{threeparttable}
\usepackage{booktabs}
\usepackage{subcaption}
\usepackage{longtable}
\usepackage{geometry}
\usepackage{graphicx}
\usepackage{hyperref}

\begin{document}
\setcitestyle{aysep={,}}

\runninghead{Alqazlan et al.}

\title{A Novel, Human-in-the-Loop Computational Grounded Theory Framework for Big Social Data}

\author{Lama Alqazlan\affilnum{1}, Zheng Fang\affilnum{1}, Michael Castelle\affilnum{2,3} and Rob Procter\affilnum{1,3}}

\affiliation{\affilnum{1}Department of Computer Science, University of Warwick, UK\\
\affilnum{2}Centre for Interdisciplinary Methodologies, University of Warwick, UK
\affilnum{3}The Alan Turing Institute, London, UK}
\email{{Lama.alqazlan|M.Castelle.1|Rob.Procter}@warwick.ac.uk,\\Lamaalqazlan@gmail.com}

\begin{abstract}
The availability of big data has significantly influenced the possibilities and methodological choices for conducting large-scale behavioural and social science research. In the context of qualitative data analysis, a major challenge is that conventional methods require intensive manual labour and are often impractical to apply to large datasets. One effective way to address this issue is by integrating emerging computational methods to overcome scalability limitations. However, a critical concern for researchers is the trustworthiness of results when Machine Learning (ML) and Natural Language Processing (NLP) tools are used to analyse such data. We argue that confidence in the credibility and robustness of results depends on adopting a 'human-in-the-loop' methodology that is able to provide researchers with control over the analytical process, while retaining the benefits of using ML and NLP. With this in mind, we propose a novel methodological framework for Computational Grounded Theory (CGT) that supports the analysis of large qualitative datasets, while maintaining the rigour of established Grounded Theory (GT) methodologies. To illustrate the framework's value, we present the results of testing it on a dataset collected from Reddit in a study aimed at understanding tutors’ experiences in the gig economy.
\end{abstract}


\maketitle

\section{Introduction and Background}
Recent advances in ML and NLP models have significantly enhanced their performance in tasks such as question-answering and text summarisation. Sometimes, these models even outperform humans \citep{devlin2018bert, he2020deberta}. This, coupled with the proliferation of big data facilitated by the widespread adoption of digital technologies, has led to an abundance of digital traces that can be collected from various sources and analysed to gain insights into human behaviour \citep{salganik2019bit,lazer2009computational}. It enables researchers to investigate new phenomena and reach a new or wider range of research subjects \citep{agarwal2014big,johnson2019revisiting}. Yet, while advanced ML and NLP are superior for some tasks, human researchers possess qualities such as socially embedded sense-making, reasoning and contextual awareness \citep{legg2007collection} that isolated and individual models trained solely on tokenised input cannot fully replace. Although recent developments have improved context handling, they still fall short of capturing the nuanced understanding inherent to human cognition \citep{wang2024survey}. Therefore, there is a need to combine the strengths of both so they complement one another when analysing large qualitative datasets \citep{herrmann2020socio,peeters2021hybrid}.

Qualitative research methods such as GT \citep{glaser1967discovery} are widely used in the humanities and social sciences. Hand-coding is central to these methods, which involve systematic data segregation, annotation, grouping and linking. The goal is to examine unstructured text to identify patterns and acquire a thorough understanding of the data \citep{chen2018using,rietz2020towards}. However, since such a procedure requires considerable manual labour and hours of dedicated effort, analysing big datasets reliably can be extremely laborious, if not impossible \citep{crowston2012using,muller2016utilizing}. To address this, some researchers use strategies such as subsampling \citep{attard2012thematic, kazmer2014distributed}, but this may fail to take advantage of the richness of the full dataset as a substantial portion of it will remain unexplored \citep{chen2018using, jiang2021supporting}. This limitation has prompted researchers to examine ways of using computational techniques to help minimise human coding efforts. In \citet{baumer2017comparing}, two researchers independently applied Topic Modelling (TM) and GT to the same dataset and compared their results. They found that many GT codes were well-represented in TM results; however, the latter had a lower level of abstraction, comparable to middle-level GT codes. Therefore, they concluded that these approaches work most effectively when combined. Similarly, \citet{bakharia2019equivalence,crowston2012using,ruis2021advances} argue that while TM may automate parts of the coding process, more work is needed to acquire a deeper qualitative insight into the social phenomena under study. The effort to achieve this has given rise to what is known as Computational Grounded Theory (CGT). Researchers suggest it is a promising methodology, but some reservations remain \citep{procter2022developing,gorra2019keep}.

\begin{table*}[t]
\centering
\footnotesize
\begin{tabular}{|p{2.5cm}|p{2.5cm}|p{2.5cm}|p{2.5cm}|p{2.5cm}|}

\hline
\textbf{Aspect of Framework} & \textbf{\citet{nelson2020computational}} & \textbf{\citet{odacioglu2022combining}} & \textbf{\citet{mangio2023s}} & \textbf{\citet{yu2019leverage}} \\
\hline
TM used & STM & LDA & STM & LDA \\
\hline
Validation of TM Topics & Not applied & Not applied & Not applied & Not applied \\
\hline
Evaluation of model's quality & top-N terms & Not applied & top-N terms and documents & Not applied \\
\hline
Qualitative Interpretation & Minimal & Moderate & Extensive & Not applied \\
\hline
Constant Comparison & Moderate & Moderate & Moderate & Minimal \\
\hline
Applying Memo-writing & Not explicitly stated & Explicitly stated & Not explicitly stated & Not explicitly stated \\
\hline
Theoretical sampling & Not applied & Not applied & Not applied & Not applied \\
\hline
Reliance on Computational Tools & Moderate & Moderate & Moderate & Heavily reliant \\
\hline
\end{tabular}
\caption{Summary of the discussed CGT Frameworks}
\label{cgt_frameworks}
\end{table*}

Several attempts have been made to develop CGT frameworks, with the most prominent being proposed by \citet{nelson2020computational} and subsequently quite widely adopted  \citep{williams2022pseudonymous,ojo2021public}. The framework has three steps that start by detecting patterns in the data using Structural TM (STM). The highest-weighted (top-N) terms (i.e., words) per topic are used to examine the model to determine if it produces semantically coherent topics. Then pattern refinement, where the researcher returns to the classified data to add interpretation through qualitative deep reading of the documents in each TM topic. Finally, the pattern confirmation step, which includes deciding which computational techniques are most suitable for verifying patterns found in the data against the complete dataset. \citet{odacioglu2022combining} propose a 4-step CGT framework, started by applying Latent Dirichlet Allocation (LDA) TM, followed by expert coding where topics are labelled based on associated terms and rated their relevance to the study. The researchers also engage in applying the codes and writing analytical memos. The third step, focus coding, involves understanding codes, identifying patterns, and seeking similarities between them by reading the associated top-N documents. Then, after constantly comparing codes and discovering links, categories are formed. The final step is theory building, which entails aggregating core categories into one higher-level theme that forms a theory.

\citet{mangio2023s}'s framework started with experts assigning topic labels to STM topics based on examining the top-N terms and on the qualitative interpretation of the top-N documents per topic. They further evaluated the performance of the model by having two researchers manually review a sample of documents for each topic to assess the model. Then, following extensive discussions, they grouped the topics into more comprehensive higher-level categories based on content similarity. The researchers then utilised the STM for validation that enabled them to analyse how topics changed over time and compare these changes with real-world events. Furthermore, two authors ‘internally validated‘ topics by manually coding a sample of documents per topic, ensuring that the model discriminated correctly. In the final step, an interpretative GT analysis of the topics is conducted. A hand-picked sample of the most representative top-N documents is independently hand-coded in three separate stages, including coding the raw data from topics, secondly, they arrange the resulting codes from first to second-order codes, which helps to provide more theoretical insights and in the final coding step a further aggregation of concepts is identified and classified into overarching categories.

While these earlier frameworks have clearly attempted to apply GT principles when utilising text mining techniques, some have oversimplified the process by over-relying on the computational tools results for theory development. \citet{yu2019leverage}’s framework is an example, which started with applying LDA and then labelling the topics based on reading the top-N terms. They viewed LDA topical terms as open coding, and LDA topic labels as axial coding. The LDAvis tool \citep{sievert2014ldavis} was then used to visualise LDA topics and they found that topics comprised three independent areas based on the topics’ relationships; therefore, they considered these areas as different categories. By further abstracting and summarising the topical terms contained in these categories, selective codes were obtained, each corresponding to a category.

In summary, the limitations in these frameworks include the absence of an initial stage of data exploration, which is critical given the amount and complexity of the data a social scientist will deal with in big data research. Additionally, there were no measures taken to externally verify the validity of the TM models’ results. Aside from this, and more importantly, certain frameworks failed to apply some of GT’s core principles, which is critical when claiming to utilise the methodology \citep{birks2013grounded}. These include not appropriately implementing the constant comparison analyses method, and lack of in-depth engagement with the data, as well as inadequately addressing the procedures for coding and grouping codes to arrive at higher-level conceptual categories that will lead to the emergence of the theory. Finally, within all the existing frameworks, there is a noticeable absence of applying theoretical sampling in big data research, a crucial step to fill in the gaps in the discovered core categories and contribute to the quality of analysis. In Table~\ref{cgt_frameworks}, a summary of the discussed CGT frameworks is presented.

\subsection{Grounded Theory}
GT is an inductive methodology for analysing qualitative data, which was originally developed by \citet{glaser1967discovery}. It was introduced to address a limitation on existing methods of social research, at that time, which primarily focused on theory verification, while the initial step of discovering relevant concepts and hypotheses inductively from data is usually overlooked. GT is ``simply a set of integrated conceptual hypotheses systematically generated to produce an inductive theory about a substantive area'' \citep{glaser2004remodeling} (p.3). This theory reveals the main concern of the research subjects and how they resolve it,  i.e. their behavioural response. It is called ‘grounded’ because it is grounded in the data: the research subjects’ own explanations or interpretations \citep{corbin1990grounded}.

The analysis begins with open coding in which the researcher codes the data line-by-line, trying to conceptualise the patterns they find, since ``the essential relationship between data and theory is a conceptual code'' \citep{glaser2004remodeling} (p.12). Initial codes are comparative and provisional \citep{charmaz2006constructing} and it is only through constant comparison --- the process of comparing incidents (i.e. the empirical data, the indicators of a category or code) with incidents, incidents with codes, and codes with codes --- categories followed by higher-level categories start to appear \citep{charmaz2006constructing}. A category represents a group of related codes that share common characteristics or patterns. These codes (or sub-patterns) consider the ‘properties’ or ‘dimensions’ of that category \citep{glaser2002conceptualization}. Writing analytical memos is another continuous process to elevate data to concepts and an important component of quality \citep{glaser2004remodeling}. As the constant comparison continues, core categories begin to emerge. A core category is considered central and may represent the main concern; frequently appears in the data and is meaningfully related to other categories \citep{glaser1967discovery}. Then, selective coding, which entitled limiting subsequent data collection and coding to variables related only to core categories \citep{glaser2004remodeling}.

As the analysis progresses, the researcher begins to identify gaps for which theoretical sampling is required to saturate the emerging core categories and provide the remaining necessary insights; here they must make decisions about which sources of data will meet the analytical needs \citep{birks2015grounded}. Only core categories with the most explanatory power should be saturated \citep{glaser1967discovery}. Then, the final stage of theoretical coding begins, which is defined by \citet{glaser1978theoretical} as conceptualising ``how the substantive codes may relate to each other as hypotheses to be integrated into a theory'' (p.72).

\section{Enhancing Trustworthiness in CGT Research}
As big data provides new opportunities for social research, establishing trust in the research tools and methodology and thus in the results is of paramount importance. Some conventional researchers have raised concerns about the trustworthiness of ML models in conducting qualitative analyses \citep{eickhoff2017topic,jiang2021supporting,drouhard2017aeonium, gao2023feasibility}, and the credibility of results due to an over-reliance on computational tools \citep{nguyen2021establishing, ramage2009topic, dellermann2019hybrid}. This can be attributed to some weak examples of utilising these tools to explain and build theories about human behaviour without grounding them in social scientific foundations and accepting the results without proper validation. Therefore, we argue that when proposing frameworks to automate GT for big data analysis, trust can first be established by assuring social scientists that the core principles of their well-established methodology are applied and taken into account to maintain its rigour.

Furthermore, trustworthiness, defined as confidence in the quality of a research study, including its methods, data, and interpretation \citep{connelly2016trustworthiness}, is vital for ensuring credibility \citep{shenton2004strategies,lincoln1985naturalistic,nguyen2021establishing}. Here, trust in computational methods must be assessed rather than assumed, given that high statistical performance in validating and evaluating ML models does not necessarily imply that their results are interpretable or practically effective \citep{chang2009reading,grimmer2011general}. Thus, many approaches have been proposed to incorporate human validation and evaluation \citep{zhang2025can,elangovan2024considers,ying2022topics,doogan2021topic,mimno2011optimizing,chang2009reading}. Within the context of CGT, it is important to ensure that the incorporated techniques effectively replicate the role of the researcher in big data analysis. We argue that when using TM to automate the coding process, researchers should externally validate these topics to ensure the model’s ability to detect reliable patterns in the data, thereby the reliability of the automated coding process. Researchers then play an important role in evaluating the quality of the chosen models’ outputs. As \citet{grimmer2021machine} argues, rather than fully relying on fitting statistics, it is important to obtain feedback from human researchers about the quality of the generated topics.

Additionally, the use of computational models in CGT can significantly contribute to fostering trust in the resulting GT by enhancing objectivity and reducing human bias \citep{shenton2004strategies}. This approach aligns with the concept of 'mechanical objectivity' as described by \citet{porter1997trust} and \citet{daston2021objectivity}, emphasising reliance on machines and the trust in numbers to produce unbiased results. At the same time, trust in the developed GT is further reinforced by preserving the researcher’s role in qualitatively interpreting the model’s outputs to derive meaningful theoretical insights. This also reflects \citet{daston2021objectivity}’s notion of 'trained judgment' where objectivity arises from the interplay between mechanical processes and human expertise \citep{benbouzid2023fairness}. Personal biases in the interpretation stage can be further mitigated by adhering to the constant comparison process \citep{glaser1967discovery}, central to conventional GT. Preserving the researcher’s interpretative role ensures that automation does not diminish the value of qualitative insights but instead enhances both trust and the researcher's ability to engage with big datasets. This is particularly important for elevating the conceptual depth of codes generated by TM and for facilitating the development of higher-level core categories—crucial steps in constructing a GT.

This active researcher engagement for achieving this balance is commonly referred to as the ‘human-in-the-loop’ (HITL) approach. In this approach, computational tools are used to support the analysis, while the researcher retains an active role throughout the process \citep{kim2018human}. Adopting HITL enables social scientists to address issues of trust and credibility when using ML methods by allowing them to intervene in validating, evaluating, and interpreting computationally generated results. This also ensures that researchers maintain a sense of closeness and control over their qualitative analyses, which might otherwise be lost in fully automated processes \citep{jiang2021supporting}.

Thus, this paper introduces a novel methodological framework for CGT that incorporates the HITL approach and addresses the limitations of existing CGT frameworks to enhance trust. Unlike researchers who emphasise algorithmic failure and 'data friction' as a means to inspire creative insights and identify significant cases for interpretation \citep{madsen2023friction,munk2022thick,rettberg2022algorithmic}, our framework takes a different stance. The primary objective is to ensure that the CGT framework and its associated techniques, can automate key aspects of the traditional methodology, making it scalable and suitable for big data contexts without compromising the foundational principles of GT.
 
\section{Methodology}
\subsection{Computational Grounded Theory: A New Approach}

\begin{figure*}[t]
\center
\includegraphics[width=5in]{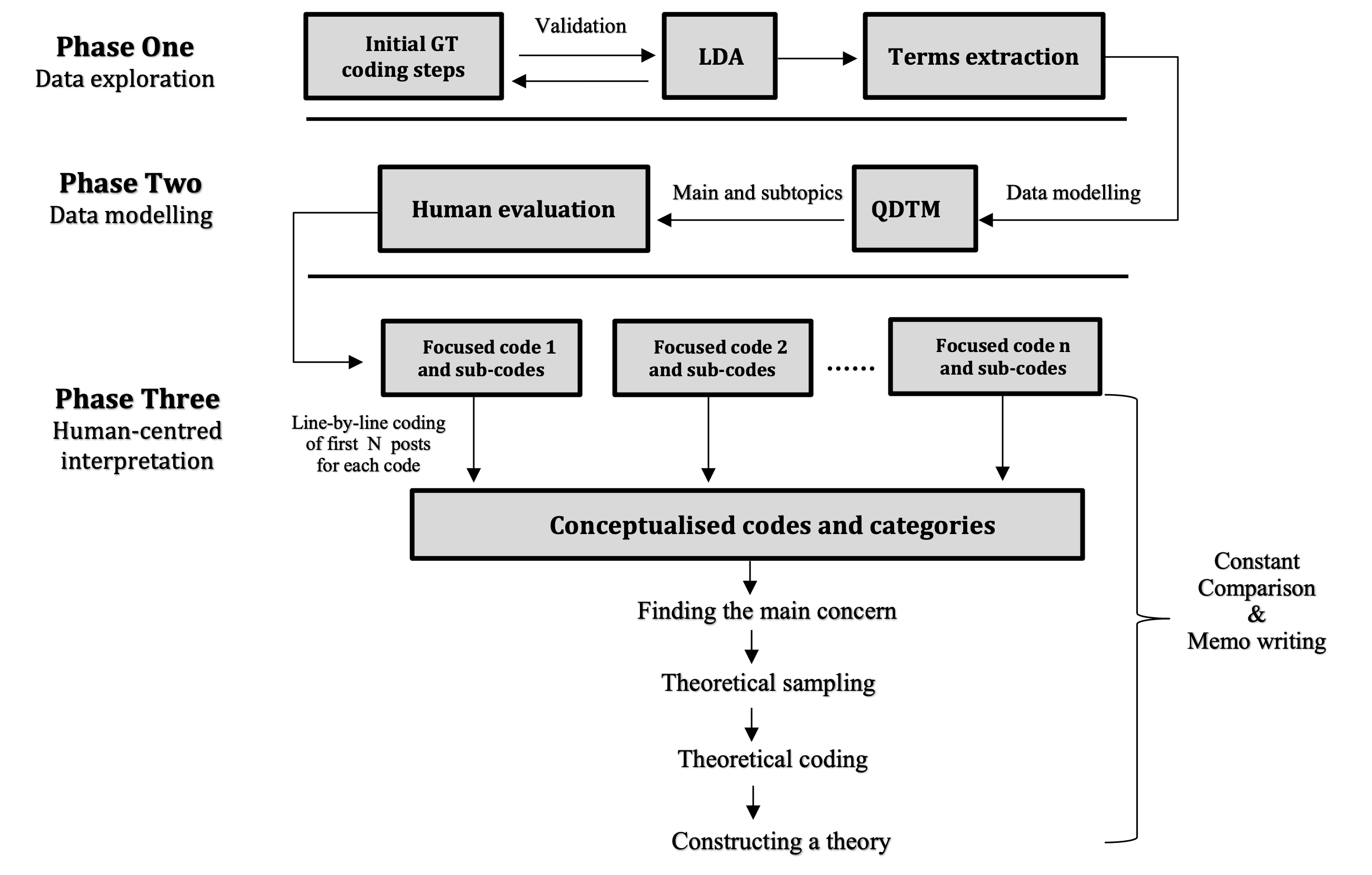}
\caption{The CGT Framework Chart}
\label{fig:Framework}
\end{figure*}

CGT frameworks were developed as a way of dealing with large volumes of data when conducting a qualitative analysis. However, preserving the fundamental principles of traditional GT to the greatest extent possible is essential to claim to have followed the methodology \citep{birks2013grounded}. Our three-phase framework (See Figure~\ref{fig:Framework}) strives to achieve this goal. Using specific computational tools enabled us to implement and facilitate the application of other GT principles to big data research, beyond the coding process that is the main focus of existing frameworks.

HITL ensures the broader social and contextual validity of the topics discovered using unsupervised TM models. The involvement of the researcher in ongoing interaction with the results generated by computational tools enhances the reliability of the analysis. The framework also provides a detailed description of how to conduct a human-centred interpretation of the computationally produced results effectively. This will further ensure adherence to GT principles, which will lead to an improved theoretical understanding of the substantive field of study and enhance the trustworthiness of the results.

\subsection{Phase One: Data Exploration}
In this phase, the researcher begins by familiarising themselves with their dataset. Given that in big data research researchers do not collect the data by engaging directly with the research subjects, gaining an understanding of the data, its structure and complexity is crucial. While \citet{goni2022experiential} suggests reading subsets of data for this purpose, we propose coding a random subset of the data using GT's initial coding steps. This will result in a list of codes with precise descriptions of the content of each one. These resulting codes should be utilised to validate the patterns that will be explored computationally using unsupervised ML methods (i.e. ones that learn without training data) when exploring a larger volume of data in the next step. This systematic approach is important to provide a valuable method for validation. Even though the original GT methodology encourages coming to the data with a 'blank slate', i.e. not bringing any assumptions about the data into the coding process, here in CGT we mainly depend on ML models to code the data, which necessitates ensuring that these substitute for human researchers effectively and appropriately. Notably, previous frameworks such as \citet{nelson2020computational} and \citet{odacioglu2022combining} did not incorporate this step.

Next, using LDA TM (See Figure~\ref{fig:Framework}), the entire dataset is explored to identify the main patterns. Upon applying LDA, each topic should be represented by a list of the top-20 terms and top-N documents (N is decided by the researcher based on the nature of their data). To label topics, we emphasise the necessity of reading topical documents, since close reading can provide better qualitative insights into the topic’s content \citep{brookes2019utility}. A similar approach is taken by \citet{mangio2023s} and \citet{nelson2020computational} while others only used topical terms to label topics, which was proven to not reflect topic content accurately. Following this, HITL validation is crucial to ensure that the TM output effectively replaces human reading and judgments on a large dataset, avoiding the blind use of unsupervised and black-boxed ML algorithms \citep{dimaggio2015adapting,grimmer2013text,korenvcic2018document}. Thus, a concurrent validation \citep{boussalis2016text} is used, where the researcher compares LDA topics with the generated GT codes. Due to LDA scalability and the size of the data included in this step, new topics might be discovered, which is to be expected and should not impact the validity of the analysis. Here, a high level of alignment between GT and LDA results will ensure validity. The similarity between the two methods is what enable validation, as both are exploratory, data-driven, and iterative \citep{muller2016machine}. Then, the resulting GT codes from step one and the LDA topics will both be considered, this includes not only the topics that overlap but also those unique to either GT or LDA, which is essential to fully leverage the power of ML models in uncovering patterns in big data.

One of the advantages offered by LDA is the availability of terms associated with each identified topic, which is an important resource that our framework utilises. In the term extraction step, the researcher refines and curates the set of terms for each of the previously labelled LDA topics that will form the basis for the data modelling phase. In essence, the term extraction step serves as a critical link between topic identification and modelling. For topics identified in GT and LDA, or only in LDA, the top-20 terms were selected. Any terms that are deemed irrelevant to the topic being discussed should be excluded from the list based on the researcher’s understanding of the topic. This process ensures that the terms associated with each topic reflect the topic content. Secondly, terms should be proposed for topics that appeared only in GT.

\subsection{Phase Two: Data Modelling}
After identifying and validating the main topics of discussion, the Query-Driven Topic Model (QDTM) \citep{fang2021query} is used. It is a semi-supervised TM approach designed to return topics relevant to queries input by the researcher in a tree-like structure. Here, the researcher inputs the terms extracted from the previous step, a set of query terms for each topic, into the model. QDTM then employs term expansion techniques — Frequency-Based Extraction, KL-Divergence-Based Extraction (KLD), and the Relevance Model with Word Embeddings — on the input queries, producing a set of concept terms to enhance topic modelling. These techniques focus on retrieval and relation rather than generation in identifying terms. Then, terms are fed into a framework based on a variation of a Hierarchical Dirichlet Process (HDP) to form main topics and a second layer of subtopics, which is considered an effective way to organise and navigate large-scale data \citep{dumais2000hierarchical,johnson1967hierarchical}. Furthermore, the HDP is a nonparametric Bayesian model that automatically infers the number of topics in a corpus \citep{teh2004sharing}.

In CGT, this hierarchically-structured TM is particularly helpful in ensuring the effective application of GT principles and maximises the researcher’s ability to extract the full analytical value from large datasets. First, since researchers will be provided with groups of related main and sub-topics together; this will greatly support the intermediate coding steps \citep{birks2015grounded} and allow for more effective application of the ongoing process of constant comparison during analysis. It will also enable inductive reasoning and a focused examination of related documents \citep{fang2021query}. Second, in traditional GT research settings, it is the researcher’s responsibility to revisit the field and collect specific data when encountering a code or category they do not fully comprehend. In the proposed framework, QDTM automates or partially automates the saturation process. In most cases, only terms automatically generated by LDA are used, with no additional terms added to reduce human bias. Thus, the input queries are expanded into a comprehensive keyword list, making the output topics more relevant. The term expansion techniques employed are highly valuable as they re-examine the complete dataset and identify various variations and dimensions within each main topic \citep{carlsen2022computational}. By revisiting the corpus to capture overlooked aspects — such as low-frequency terms or niche topics — that were not initially captured due to their infrequent occurrence or any other factors, QDTM presents the researcher with enough variations and cases for each main topic. This will significantly reduce the necessity for researchers to engage in ‘extensive’ theoretical sampling to saturate their understanding of the identified codes or categories. Despite that, there may still be instances where such sampling might be necessary. Finally, it is critical that the number of sub-topics is automatically determined; in real-world large datasets topics are not evenly distributed, some are more frequent and have a wider range of variation than others. This feature might help researchers gain insights regarding the main concerns of research subjects and the core categories, since one of the indicators is its frequency of occurrence in the data \citep{glaser1967discovery}. 

These distinctive features of QDTM go beyond automating the coding process in large datasets. Instead, they actively support the application of other core principles of GT in big data research, making it a more suitable choice compared to single-layered TM models used in existing CGT frameworks.

\subsection{Human Evaluation of QDTM Topics}
The aim of this step is to ensure that the QDTM is capable of generating high-quality topics. This is critical as it will enhance the researcher’s ability to interpret these topics more efficiently and confidently at a later stage in the framework. To perform the evaluation tasks reliably, two or more annotators must independently annotate the data \citep{pustejovsky2012natural}. This step further demonstrates the active involvement of the HITL in our framework and how this can contribute to increased trust in the final results. We emphasise the importance of evaluating topics' intrinsic semantic quality. This is typically assessed using statistical measures, such as perplexity, however, negative correlations between these measures and human judgments of topic quality have been observed \citep{chang2009reading}. Hence, we turn to HITL evaluation as an alternative -- and superior --- method. Typically, tasks for human evaluators (annotators) might include examining the quality of top-N terms for each topic \citep{mimno2011optimizing}, or to identify the intruder within these terms \citep{chang2009reading}. However, despite the popularity of measuring coherence based on topical terms, this approach is criticised for only partially reflecting topic content and quality \citep{brookes2019utility,korenvcic2018document,grimmer2011general}. 

In our framework, human evaluation is used to assess QDTM results, which is primarily based on topic documents. However, a list of terms per topic is provided to be used in specific cases. To effectively evaluate and annotate the QDTM's topics, four distinct tasks should be undertaken. Firstly, rate the quality 'coherence' of each topic, then identify the issues with lower quality topics, and then label the topics. Furthermore, since topics are hierarchically classified, this will only partly reflect output quality \citep{belyy2018quality, koltcov2021analysis}, the relationship between main and subtopics should also evaluated. This step will enhance the researcher's ability to interpret these topics more efficiently and confidently at a later stage. In Appendix A the complete annotation guidelines are presented. 

\subsection{Phase Three: Human-Centred Interpretation}
This final phase of the proposed CGT framework is dedicated to the in-depth, interpretative analysis of the topics produced by the QDTM. Based on the human evaluation conducted in the previous phase, the researcher will be presented with N groups of related topics, each representing a main topic and a number of subtopics. In the framework, the main topics are considered “focused codes”, while the subtopics are considered “sub-codes” (See Figure~\ref{fig:Framework}). The labels produced by the annotators for each code are considered initial labels.

The interpretative analysis is started by coding the representative documents for each topic line-by-line. This process of 'hand coding' enables researchers to “zoom in” on the studied phenomena for appropriate theoretical interpretations \citep{aranda2021big,mangio2023s} and will increase confidence that nothing has been overlooked \citep{glaser1967discovery, glaser2004remodeling}. As instructed by GT's originators, hand-coding is vital to stimulating ideas, writing analytical memos, and properly applying comparative analysis to create higher-level categories. Here, since computational tools are used to classify the data and then the annotators to evaluate the resulting topics (codes), researchers are required to do their own coding. A similar approach is taken by \citet{mangio2023s}, whereas in \citet{nelson2020computational,odacioglu2022combining}, interpretations are added to the analysis through a simple reading of the documents. 

At this point, the size of documents will influence the plausible number of documents that can be manually coded and analysed \citep{amaya2021new}. Researchers should try to achieve a balance between the need for thorough analysis to draw meaningful insights and the practical constraints in terms of the resources and time available for manual analysis.

The analysis begins by coding the top-N posts of the first focused code. Constant comparison is applied here by comparing incidents in data against the code label and with each other. During this process, more sub-codes might be created and the researcher should raise the conceptual level of annotators' labels since they are descriptive of topic content. After analysing the first focused code, the same process is applied to the first sub-code from the same group. The resulting codes should then be compared; the researcher should look for patterns, links, or any relationship between the focused code and the sub-code. This may reveal that one code (whether focused or sub-code) is part of the other or create a higher-level category that connects them. Here, the researcher should allow the analysis to take its course and ensure that the memos are written on a regular basis. Similar processes is applied for each of the following sub-codes until the analysis of the first group is completed. The researcher then moves on to the next group, until all are analysed.

The resulting codes and categories from each group are then compared to find relationships and to construct higher-level categories. At this point, the core categories and the main concern of the subjects under study should be determined and confirmed. Then, the researcher should engage in theoretical sampling to saturate gaps found in core categories (see next section). At this stage, the theoretical coding step commences, which is the process of defining possible relationships between the developed fully saturated core categories. \citet{glaser1978theoretical} explains that ``Theoretical codes implicitly conceptualise how the substantive codes will relate to each other as interrelated multivariate hypotheses in accounting for resolving the main concern'' (p.163). As opposed to the coding in the previous steps, which is data-driven, here we deal with ideas and perspectives which will help the researcher integrate categories and data into a theory.

\subsection{Supporting Theoretical Sampling}
Theoretical sampling significantly improves the quality of the analysis by allowing researchers to collect additional selective data to fill in emerging categories and address questions raised during the analysis \citep{charmaz2012power}, and is an often-overlooked step in big data research. In conventional GT studies, it is typically performed by following up with participants, observing in new settings, etc. However, in big data research, returning to data collection is not always feasible, for example, when the complete dataset is collected before analysis begins \citep{birks2013grounded}. According to \citet{glaser1967discovery}, theoretical sampling is open and flexible: ''Theoretical sampling for saturation of a category allows a multi-faceted investigation, in which there are no limits to the techniques of data collection, the way they are used, or the types of data acquired'' (p.65). Therefore, in the proposed CGT framework, we argue that theoretical sampling can be implemented in big data research by utilising different computational techniques on the same sufficiently large and rich dataset. For example, the research might use sentiment or time series analysis, depending on the questions raised during the analysis, to help them gain completely different but needed results. This means conducting a multi-method analysis of the same dataset, previously analysed using TM. Here, researchers have to determine the most suitable NLP technique based on their needs, meaning that sampling methods will vary across different studies. Therefore, as also asserted by \citet{charmaz2014constructing}, theoretical sampling is a strategy that researchers employ and adapt to fit their specific study needs, not a standard data collection method.

\section{Case Study}
To illustrate the implementation and usefulness of the proposed framework, a case study of gig economy tutors was conducted, with the ultimate goal of developing a substantive theory that could help explain their experiences.

Global economy digitalisation is associated with the rise of the gig economy, in which long-term employment is replaced by one-off transactions based on performing single tasks. This non-standard work, mediated through online platforms, has permitted flexible work arrangements and links independent workers with customers globally. It provides greater job autonomy and expanded employment opportunities \citep{clark2021gig}. Despite that, the shift from market for jobs to market for tasks has also resulted in the loss of protections and rights typically associated with long-term employment \citep{drahokoupil2021modern}. It is linked with job insecurity, income instability and low wages. This leads workers to work longer hours to achieve their earnings goals \citep{clark2021gig,duggan2022boundaryless}. Moreover, there is an ethical concern arising due to, for example, a lack of transparency in platform operations \citep{jarrahi2019algorithmic}, which may create power imbalances between workers and platforms \citep{koutsimpogiorgos2020conceptualizing}. 

The nature of tasks performed in the gig economy has a significant impact on the challenges encountered and the overall experience of workers. This case study examined the experiences of an under-studied group of workers, tutors, where one-to-one online educational sessions are conducted, making them distinctive from other common forms of gig work e.g. in food delivery, transportation services, or isolated knowledge-work tasks.

\subsection{Case study data}
CGT frameworks are generalisable to any source of textual big data. In our case study, we collected data from Reddit for illustration purposes and to test the framework. We examined Reddit and identified 18 relevant subreddits (see Table~\ref{Subreddits}) in which our target population shares their experiences and asks questions. Using the Reddit API Wrapper, approximately 52K posts and comments were retrieved. For training the TM models, pre-processing of the data was performed, including the removal of stopwords, punctuation, emojis, URLs, lowercasing, lemmatisation and tokenisation. This resulted in a vocabulary size of 7,491. Ethical approval was obtained from the Biomedical and Scientific Research Ethics Committee (BSREC) at the University of Warwick before collecting the data.

\begin{table}[h]
    \centering
    \footnotesize
    \begin{subtable}[t]{0.45\textwidth}
        \centering
        \caption{Platform-specific Subreddits}
        \begin{tabular}{@{}ll@{}}
            \toprule
            \textbf{Subreddit name}   & \textbf{Number of posts} \\ 
            \midrule
            Vipkid          & 11457           \\ 
            DaDaABC         & 10857           \\ 
            Cambly          & 8082            \\ 
            iTutor          & 6057            \\ 
            iTalki          & 1770            \\ 
            MagicEars       & 1300            \\ 
            Qkids           & 1291            \\ 
            Gogokid         & 680             \\ 
            ZebraEnglish    & 290             \\ 
            Tombac          & 154             \\ 
            Preply          & 144             \\ 
            GoGoKidTeach    & 129             \\ 
            Palfish         & 44              \\ 
            \bottomrule
        \end{tabular} 
    \end{subtable}
    
    \vspace{1cm} 
    
    \begin{subtable}[t]{0.45\textwidth}
        \centering
        \caption{Subject-specific and other related Subreddits}
        \begin{tabular}{@{}ll@{}}
            \toprule
            \textbf{Subreddit name}           & \textbf{Number of posts} \\ 
            \midrule
            OnlineESLTeaching       & 9483            \\ 
            Online\_tefl            & 2260            \\ 
            OnlineESLjobs           & 119             \\ 
            TeachEnglishOnline      & 101             \\ 
            Teachingonline          & 49              \\ 
            \bottomrule
        \end{tabular}
    \end{subtable} 
        
    \caption{The subreddits included in the study, along with the number of posts.}
    \label{Subreddits}
\end{table}

\subsection{Phase One}
To apply the first step to the case study, given the fact that our dataset consists of 18 subreddits, two relatively small subreddits were randomly selected, namely GoGoKidTeach and Palfish. This selection resulted in about 160 posts. The analysis process started by coding the data line-by-line, and since GT is a comparative and iterative process, the resulting codes were compared to one another, leading to the development of higher-level codes that group related codes together. After employing this step to this subset of the data, a total of 15 codes were identified (see Table~\ref{GT_initial}). Due to their abstract nature, two of these codes, namely `Sharing experiences and feelings' and `Seeking and providing help and advice' were excluded from the comparison with LDA in the next step. This decision was made because these are contextual and primarily reflect the underlying purpose of the posts, and LDA as a statistical model is not expected to model them, which is designed to identify more concrete and topic-based patterns, rather than abstract themes tied to emotional expressions or interpersonal interactions.

Regarding the LDA process, two models with 13 and 17 K-values were tested \citep{alqazlan2021using}. The methods used to determine the value of K is firstly by following the approach of \citet{quinn2010analyze,grimmer2010representational} to empirically examine the performance of LDA models with different numbers of topics. Then, due to its time-consuming nature, we used the Tmtoolkit \renewcommand{\thefootnote}{\arabic{footnote}}\footnote{{https://tmtoolkit.readthedocs.io/en/latest/topic\textunderscore modeling.html\#Computing-topic-models-in-parallel}} Python package to compute and evaluate several models in parallel using state-of-the-art theoretical approaches, and based on this 17 topics were eventually selected. Evaluating the quality of the topics and labelling them was performed by examining the top-5 documents.

For validation, by considering both LDA models (with 13 and 17 topics), it was found that they were collectively able to detect 12 topics that were identified by GT analysis. Yet, both models failed to model one topic –- 'Covid-19-related discussions' –- that was present in GT codes. Conversely, only one topic -- 'Bank transfers and transaction fees' -- was clearly modelled in both LDA models but did not appear in GT codes. Therefore, since the aim of this step was to validate and further explore the data, and only one new topic was found even after the number of topics was increased to 17, the aim of this step was fulfilled, and it was not necessary to examine further models.  Here, the obtained codes and topics will both be considered, making the final output of this process 14 topics that will be taken into account in the next phase. In Appendix B, the term extraction results needed for applying QDTM are presented. 

\begin{table}[t]
\footnotesize
\centering
\begin{tabular}{l|l}
\hline
\textbf{Code no.} & \textbf{Label} \\
\hline
Code 1 & Job requirements \\
Code 2 & Hiring process \\
Code 3 & New contracts\\
Code 4 & How tutors consider the job \\
Code 5 & Reasons to join or leave a platform \\
Code 6 & The class and the students \\
Code 7 & Teaching material and methods \\
Code 8 & Bookings and working hours \\
Code 9 & The payments \\
Code 10 & The rating system \\
Code 11 & Technical issues \\
Code 12 & Miscommunication with platforms management \\
Code 13 & Covid-19-related discussions \\
Code 14* & Sharing experiences and feelings \\
Code 15* & Seeking and providing help and advice \\
\hline
\end{tabular}
\caption{List of codes derived from the initial exploration step.}
\label{GT_initial}
\end{table}

\subsection{Phase Two}
To run QDTM on the gig economy tutors dataset, the model was input with a list of terms for each of the 14 topics that were identified at the terms extraction step. Here, after applying the pre-processing steps and training the model, vocabularies with less than five document frequencies and sub-topics with very low prevalence (less than 0.2\%) were excluded. Once we were satisfied with the resulting model, we removed duplicate posts that appeared in the top-10 posts of each topic, and identical subtopics within each topic based on lower prevalence. This resulted in 76 topics (i.e. 14 main topics and 62 subtopics), each represented as top-5 posts and top-10 terms for the next step. 

The human evaluation task was conducted by three experienced journalists. The reason for choosing this group is because they had taken part in similar annotation tasks before, which helped to significantly reduce the training time. Moreover, as journalists are generally well-acquainted with many different subjects, making it easier for them to understand the data. In total, each annotator spent about 15 hours working to complete the annotation work. They followed guidelines developed by the research group through several rounds of revision and testing (See Appendix A). Overall, the annotation was conducted in two stages; first, two main topics and their subtopics were annotated (12 topics total), and the second was with the remaining 12 topics and subtopics (64 topics total). The Inter-Annotator Agreement (IAA) was calculated for each one. Four tasks were required: evaluating topic quality, identifying issues with topics, labelling topics, and evaluating relationships between main topics and subtopics. Finally, the criteria for exclusion of topics is that main or subtopics are found to be incoherent, subtopics that are unrelated to their main topics, and topics where all three annotators disagreed on one or more tasks.

To measure the IAA of human evaluation of topics, Fleiss' kappa \citep{fleiss1971measuring} was used (see Table~\ref{IAA}). According to \citet{landis1977measurement}, all obtained scores are considered 'fair agreement'. However, we found that in 97\% of the annotations, at least two of the three annotators agreed on all tasks. This indicated that the results are acceptable \citep{feinstein1990high}, and it was not necessary to take further steps to increase agreement. Moreover, task dependency may have contributed strongly to the obtained results, as disagreements in task one will usually result in disagreements in tasks two and four. For final annotation decisions, we used a simple majority voting method, and in cases of complete disagreement among the annotators (only 7 annotations, see Table~\ref{IAA}), these topics were excluded. The final category counts after resolving disagreements are in Table~\ref{AfterResolving}. Based on our exclusion criteria, 21 topics (27\%) out of 76 were excluded. Therefore, a total of 55 topics will be considered in the next phase of the analysis. Of these, 14 were main topics and 41 were subtopics. 

\begin{table*}[t]
\setlength\tabcolsep{4pt} 
\small
\begin{tabular*}{\textwidth}{@{\extracolsep{\fill}} l *{6}{c} }
\toprule
\textbf{Stage} & \textbf{Task} & \makecell{\textbf{All agree}} & \makecell{\textbf{Two agree}} & \makecell{\textbf{No agreement}} & \makecell{\textbf{Total number of topics}} & \makecell{\textbf{Fleiss' kappa}} \\
\midrule
\multirow{3}{*}{\textbf{One}} & Topic coherence & 6 (50\%) & 6 (50\%) & 0 & 12 & 0.34 \\
& Issue Identification & 5 (42\%) & 7 (58\%) & 0 & 12 & 0.38 \\
& Relatedness to main topic & 4 (40\%) & 6 (60\%) & 0 & 10 (only applied to subtopics) & 0.38 \\
\midrule
\multirow{3}{*}{\textbf{Two}} & Topic coherence & 32 (42\%) & 43 (57\%) & 1 (1\%) & 76 & 0.33 \\
& Issue Identification & 24 (32\%) & 49 (64\%) & 3 (4\%) & 76 & 0.35 \\
& Relatedness to main topic & 18 (29\%) & 41 (66\%) & 3 (5\%) & 62 (only applied to subtopics) & 0.21 \\
\bottomrule
\end{tabular*}
\caption{IAA results.}
\label{IAA}
\end{table*}

\begin{table*}[t]
\setlength\tabcolsep{5pt} 
\small
\begin{tabular*}{\textwidth}{@{\extracolsep{\fill}}l*{7}{l}}
\hline
\textbf{Task One:} & \multicolumn{5}{l}{\textbf{Topic coherence}} & \textbf{Total} \\
\hline
\multirow{3}{*}{Number of topics} & Coherent & Average & \multicolumn{2}{l}{Incoherent} & No agreement & \\
& 42 & 24 & \multicolumn{2}{l}{9} & 1 & 76 \\
\hline
\textbf{Task Two:} & \multicolumn{5}{l}{\textbf{Issue identification}} & \textbf{Total} \\
\hline
\multirow{3}{*}{Number of topics} & Intruded & Chained & Random & No issue & No agreement & \\ 
& 23 & 4 & 4 & 42 & 3 & 76 \\
\hline
\textbf{Task Three:} & \multicolumn{5}{l}{\textbf{Relatedness to main topic}} & \textbf{Total} \\
\hline
\multirow{3}{*}{Number of topics} & Strongly related & Partially related & Not related & Identical & No agreement & \\ 
& 34 & 7 & 18 & 0 & 3 & 62 \\
\hline
\end{tabular*}
\caption{Final categories counts after resolving disagreement}
\label{AfterResolving}
\end{table*}

\subsection{Commencement of Phase Three Analysis}
Following the quality evaluation of the QDTM results, it was necessary to determine the length of posts to decide how many posts it would be feasible to include in the analysis. Across 55 topics, statistics of the top 10 posts revealed a minimum length of 11 tokens, an average of 165 tokens, and only 13 posts exceeded 512 tokens. Based on these, it was determined that hand-coding of 550 posts would be a manageable task. We started the analysis by examining groups of related topics, and manually coding the top-10 posts for each code in the first group. Then, we proceeded to analyse the next group, continuing this process until all 14 groups were examined. Subsequently, after identifying the specific category that necessitated theoretical sampling, we integrated the newly identified data into the coding process and compared it with other data within that category.

\begin{figure*}[t]
 \center
  \includegraphics[width=6in]{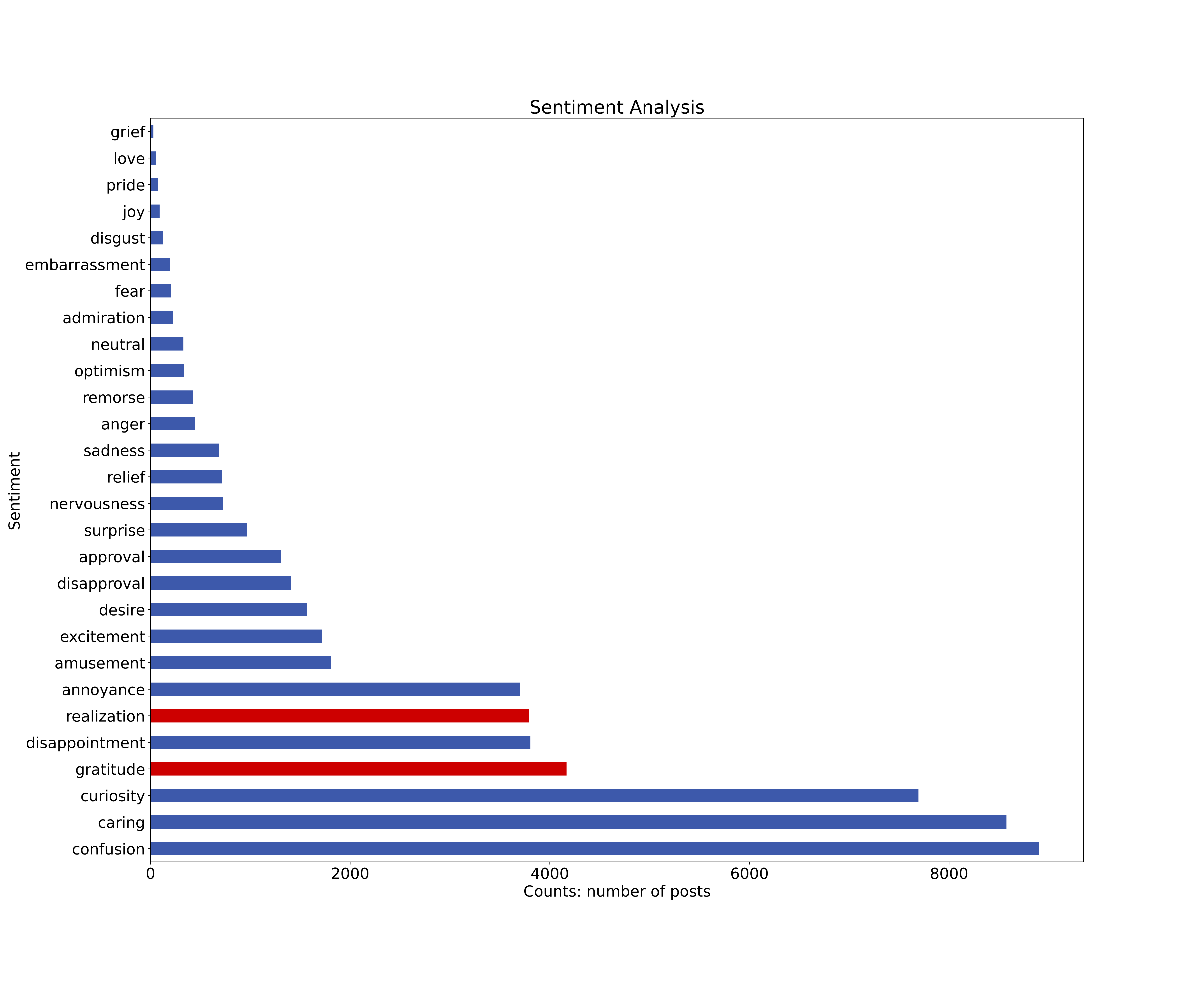}
  \caption{The frequencies of emotions in our dataset. The bars in red are the two emotions selected for theoretical sampling. Gratitude represents about 7.70\% and realization represents about 7.00\% of the data.}
  \label{TSfigure}

\end{figure*}

During the analysis, a category `Redditing' emerged, where tutors create their supportive network on Reddit; novices seek help and experts provide assistance. Yet, a question arose regarding whether novices truly benefited from this interaction. As a result, more empirical evidence is needed, which requires theoretical sampling. Here, a sentiment analysis model was identified to be useful in answering this question. The model {\it distilbert-base-uncased-go-emotions-student} \renewcommand{\thefootnote}{\arabic{footnote}}\footnote{{https://huggingface.co/joeddav/distilbert-base-uncased-go-emotions-student}} was pre-trained \citep{wolf2020transformers} on the GoEmotions dataset \citep{demszky2020goemotions}. This benchmark dataset contains 58K comments was manually annotated for 27 emotions. The generalisability of the data across domains was tested via different transfer learning experiments. After applying the model, we purposely selected ``Gratitude'' and ``Realization'' emotions (See Figure~\ref{TSfigure}) for further analysis. Then, we randomly sampled 50 posts assigned to each emotion (See Table~\ref{TSSamples}). Our analysis revealed that the data met our needs, leading to the creation of a new code, `Realising' as a result of tutors' Redditing behaviour.

\begin{table*}[h]
\begin{tabular}{
p{\dimexpr15cm-2\tabcolsep-1.13\arrayrulewidth}
|p{\dimexpr2.2cm-2\tabcolsep-1.13\arrayrulewidth}
}
\hline
\textbf{Example   of posts} & \textbf{category} \\ \hline
That's awesome, thanks. It will save me a lot of anxiety.  & Gratitude   \\
Thanks a lot! I wish that had been clear from the start.                                                                                                                                             & Gratitude   \\
That's genius! I really appreciate the tip! I'm definitely going to give it a try!                                                                                                                 & Gratitude   \\
It worked.. my reservations remain but I can now enjoy a 15 minute tea break without being interrupted by trial students calls. Many thanks, Herr!                                                   & Gratitude   \\ \hline
I totally get why and it makes sense!                                                                                                                                                                & Realization \\
Aha!   It is now clear to me what needs to be written in my ticket... Now I know what's missing!                                                                                                     & Realization \\
You have painted a very clear picture here! Brings a whole new perspective!                                                                                                                          & Realization \\
You made it all make sense in the way you wrote it! I knew most of that, but your way of expressing it really made sense to me!  What a great teacher you are, haha.                                 & Realization \\ \hline
\end{tabular}%
  \caption{Random samples from the data assigned to ``Gratitude'' and ``Realization'' emotions.}
  \label{TSSamples}
\end{table*}

Table 3 in Appendix C shows the resulting focused codes and sub-codes after line-by-line coding for each group of topics. Then, we compared codes and categories resulting from the analysis of all groups to identify relations and construct higher-level core categories, see Table 4 in Appendix C. Quotations from the data to support each of these core categories and corresponding codes can be found in Appendix D.

Following this, the final set of fully saturated core categories was organised based on patterns of connectivity in the theoretical coding step which greatly aids in the development of our GT. The section below presents the final product of the HITL CGT framework. 

\subsection{The GT of tutors' experiences in the gig economy}
The analysis revealed a grounded theory of tutors' experiences in the gig economy. The theory proposes that the underlying main concern underpinning most of tutors' discussions is `staying financially afloat' when working within `under-structured organisations' these are tutoring platforms. Tutors face significant `financial uncertainty' due to job insecurity, income instability, fluctuating demand, inadequate payments as well as difficult-to-attain incentives. Tutors’ ability to stay financially afloat is also challenged by and arises from opaque and often error-laden `technological systems'. The technological systems contributes to payment instability when they incorrectly mark tutors as absent and penalise tutors when technical issues prevent task completion. Tutors' income is also directly affected when they encounter issues arising from the booking system in events of overlapping classes, students' no-shows, and cancellations. The vague and often unfair rating system that evaluates tutor performance and influences students' hiring choices adds to concerns of payment instability. The tutors, however, are largely impotent when trying to address inaccuracies and malfunctions, since the tutoring platforms often fail to comply with the policies they set and lack effective communication channels in an apparent abuse of power. This category and its corresponding codes are among the most frequently discussed topics (see Table 3 in Appendix C, topics 3,4,5,6,7,8,9,and 13).

The inherent ambiguity in platform work, including unclear decision-making processes and poorly defined rules and procedures, results in a state of `confusion' among tutors. This code was found frequently in most of the 14 topics, as seen in Table 3 in Appendix C and later raised to a category in Table 4 in Appendix C. This confusion has a profound impact on how tutors act to resolve their concerns, as does their level of competency in navigating the technological systems. Platform work `experts' are those who have dedicated substantial time and effort to `understand' the job and the system. While `novices', who may or may not be task experts but are not familiar with platform work, need more time to reduce their confusion and establish themselves on the platform, especially as the technologies employed do not operate in their favour. Furthermore, tutors' `financial structure' will hugely affect the possibility of pursuing this job full-time: For example, only those with financial support or living in low-cost countries find it worthwhile to absorb the impact of uncertain payments. For the majority, achieving earning goals will negatively impact a tutor's work-life balance and job enjoyment (see Topic 10,  Table 3 in Appendix C). The necessity for them to generate income is what drives them to `continue tutoring' and adapt to challenges. 

Tutors will `persist' while working on tutoring platforms; it is the core category that is observed in nearly all topics and it illustrates the behaviour through which tutors addressed their main concern. They seek to establish a supportive network on Reddit and engage in `Redditting' behaviour as a response to the absence of workplace and colleague support. Within Reddit, tutors play different roles based on their competency levels: novices are the help seekers who operate in a context of confusion; experts provide explanations and solutions based on their understanding of the platforms and the business models. Tutors 'resist' challenges by empowering one another and forming bonds through the shared negative experiences and emotions. Tutors also persist in their efforts to `solve' the problems they encounter. This may involve communicating with the platforms' management despite their non-responsiveness, and educating students about the technology used on the platforms. To stay competitive, tutors develop interpersonal skills and use self-marketing by promoting their expertise and qualifications in their profiles. `Strategising' is another form of tutors' persistence. To mitigate income instability, tutors establish a base of regular students by maximising availability, working during peak hours and opening slots in advance. Teaching popular classes and being likeable are ways to raise popularity. Working mainly with loyal regular students and asking for high ratings are strategies for protecting reputation. `Multi-homing', i.e. working on multiple platforms simultaneously, is another strategy to combat an insufficient and unstable income.

With persistence, and dedicating time and effort, some tutors were able to resolve their main concern by `generating sufficient income' and achieving their earning goals. `Redditting' has a significant impact on reducing tutors' confusion and thus `realising' and gaining a better understanding of unclear aspects of the platform, based on our theoretical sampling results. Lastly, tutors who possess clarity and understanding of the gig economy business model are divided into two types. The first accept the flaws of the business model, set realistic expectations and do not rely on this job as their main source of income. The second type refuses to tolerate poor working conditions and low pay, opting instead to leave the tutoring platform. They may employ leaving strategies such as `platform hopping' in order to find better-paying platforms or they may switch to platforms that let them set their own prices. Alternatively they may simply choose to work as freelance online tutors who are not associated with any platform. This can be observed in Table 3, Appendix C, topics 14, 12, and the sub-codes in topics 10 and 5.

\section{Discussion and Conclusions}  
This paper proposes a novel framework for applying CGT, offering a practical solution for analysing large, qualitative data using ML and NLP tools. The trustworthiness and credibility of the results is ensured by adopting a HITL approach, and the rigour of the GT methodology was maintained by adhering to all its fundamental principles. The case study results demonstrate the potential of this novel approach to CGT, as well as making an important contribution to gig economy research by focusing on an understudied group.

Our framework introduced a departure from the common practice in CGT frameworks of using single-layered TM for coding big datasets by employing a hierarchically structured TM, QDTM. The researcher begins by using LDA to identify the main topics of discussion. Then, to ensure codes and categories saturation, the QDTM with its terms expansion techniques, extends the discovered (and validated) main topics by revisiting the complete corpus to search for all relevant sub-topics, ensuring that no pertinent data or aspects are overlooked. Furthermore, the tree-like structure of QDTM topics presents the researcher with groups of related topics, which makes it easier to analyse and identify relationships between them, thus facilitating the application of GT's constant comparison analysis. Additionally, QDTM's capability to model topics using query terms elevates its superiority over other hierarchical TM models, offering researchers the flexibility to formulate missing topics from LDA.

Our framework has maintained a balanced approach, leveraging computational tools as supportive aids rather than replacements for human involvement. We contend that building trust in the credibility of CGT is ensured by the active inclusion of researchers. Within our framework, researchers are engaged in each step of the analysis, starting from the discovery of topics, validation, the modelling process, evaluation, and ultimately, the interpretation. Furthermore, the researcher adheres closely to GT's core principles in the final phase of 'zooming in' on the phenomena under study to add interpretation and theoretical insights into the analysis. This helps to raise the conceptual level of the codes generated by QDTM and facilitates the construction of core categories. We argue that line-by-line coding of the data is what enables the discovery of research subjects’ main concern and helps to reveal how they address that, which are the key goals in any GT study.

Finally, while our research was carried out before Large Language Models (LLMs) became available, our CGT framework is effectively NLP technology agnostic and hence as more powerful NLP tools become available, they can easily be accommodated within it. 

\section*{Limitations}
NLP researchers may find a limitation in the framework regarding the reproducibility of findings derived from manual interpretive line-by-line coding. Unlike quantitative studies, ensuring identical analysis by two qualitative researchers is challenging. \citet{glaser2007constructivist} argue that through constant comparison, personal input decreases and data become more objective. We argue that computational tools in the framework can reduce subjectivity and bias. Algorithms classify texts the same way by keeping the parameters the same. For instance, QDTM generates comparable topics and sub-topics if another researcher employs the same terms provided in Table 2, Appendix B on the same dataset. Thus, the computational component can contribute to the reproducibility of the analysis. 

With regards to validation, from a GT perspective, the theories produced are inherently grounded in the data and offer `a valid explanation' of how research subjects address their concerns. Nevertheless, researchers should strive to maximise the validity of their analysis \citep{cohen2001research}. Our CGT framework ensured the validity of the quantitative analysis carried out in the first phase to identify the main topics. However, the final theory generated through the framework remains firmly situated within the qualitative realm and cannot be validated using common practices applied in quantitative and computational disciplines. 

Finally, we acknowledge that the implementation of our framework may be time consuming. However, each step is indispensable for the proper application of core GT principles and to ensure trust, especially with large datasets. Overall, as noted by its originators \citep{glaser2004remodeling}, GT's inductive and iterative nature is complex and demands a great deal of time and effort.

\bibliographystyle{sageh}
\bibliography{references}

\def\volumeyear{2024}

\onecolumn


\begin{sm}

\section{Appendix A: Annotation Guidelines for QDTM Topics}
\label{AppendixA} 

\subsection{Overview and task introduction:}
The purpose of this annotation task is to assess the quality of a text classification task done by a machine learning model. The topic modelling models classify big qualitative data into topics. Each topic should represent a coherent idea or area of discussion within the larger body of text. The goal is to categorise the documents into these distinct topics to make it easier to understand and analyse the overall content.

You will be presented with collections of documents (i.e posts) that have been classified into group of topics using the QDTM. Each topic will be divided into one main topic and its respective subtopics. These topics will be presented as a list of 5 posts and 10 terms that have the highest probability of appearing in that topic. Overall, the annotation tasks are to:

\begin{enumerate}
\item Evaluate the quality (coherence) of each topic, meaning the degree to which the posts in a topic are related to each other and discuss the same subject. A \textbf{subject} is a group of at least two posts referring to the same topic.\\
\item Identify the reasons for lower ratings for topics rated as average coherence and incoherent.\\
\item Provide a label for each topic.\\
\item Determine the degree to which each subtopic is related to the main topic. 
\end{enumerate}

\subsection{Annotation procedure}
You will be provided with an Excel workbook with number of spreadsheets. Each spreadsheet consists of one main topic and its respective subtopics for which you are requested to perform the following four tasks: 

\subsection{Task One: Rate the quality of topics}

\begin{enumerate}
\item Carefully read all five associated posts and identify the topic of discussion in each post.
\item Decide whether the set of posts is coherent or not and assign a coherence score on a scale of 1 to 3 as follows:

\begin{enumerate}
\item Assign a coherence score of 3 (Coherent) to the topic if you conclude that all five posts are related and discuss the same subject.   

\item Assign a coherence score of 2 (Average) to the topic if you conclude that three or four out of the five posts are related and discuss the same subject.

\item Assign a coherence score of 1 (Incoherent) to the topic if you conclude that only two out of the five posts are related and discuss the same subject or that all the posts are unrelated or random posts.
\end{enumerate}
\end{enumerate}

\textbf{Dealing with posts that discuss more than one topic:}
 Some posts can be quite long, you might find more than one topic discussed in a single post. In these circumstances, please read through the available list of topical terms to help you decide what you consider to be the main topic of discussion.  

\subsection{Task Two: Issue identification}
For the topics that have been rated as average or incoherent quality, in the Issue Identification column, please specify the ``post number/s`` that are causing the incoherence in the topic and identify the issue type by marking it on the spreadsheet as the following: 

\begin{enumerate}
\item \textbf{Chained:} applies when two subjects can be identified in the topic (including if there is a random post that does not belong to either subject). 

\item \textbf{Intruded:} applies when one common subject can be identified in 2 –- 4 of the posts in the topic and the remainder do not have any clear connection to the subject or between each other. 

\item \textbf{Random:} applies when no clear connections between any of the posts in the topic can be identified. ‘Random’ should only assigned to an incoherent topic.  

\item Please type \textbf{(N/A)} in the column for topics rated as 3 (Coherent).
\end{enumerate}

The table below illustrates the possible outcomes of the issue identification task:

\begin{longtable}[c]
{|p{0.1\columnwidth}|p{0.1\columnwidth}|p{0.2\columnwidth}|p{0.3\columnwidth}|p{0.15\columnwidth}|}
\caption{Possible outcomes of the issue identification task}
 \\ \hline
  Topic coherence & Number of related posts & Unrelated posts & Justification & Examples \\ \hline 
  \endfirsthead
  \hline
  Topic coherence & Number of related posts & Unrelated posts & Justification & Examples \\ \hline 
  \endhead
  \hline
  \endfoot
  Coherent & (5) & - & No issue & N/A \\ \hline 
  Average & (4) & (1) & Intruded – first 4 posts about one subject, Post5 different topic & Post5 - Intruded \\ \hline 
  Average & (3),(2) & -- & Chained –- two subjects discussed & Post4, Post5 -- Chained \\ \hline 
  Average & (3) & (1),(1) & Intruded –- 3 posts one subject, others different topics & Post1, Post2 -- Intruded \\ \hline 
  Incoherent & (2),(2)* & (1) & Chained –- two subjects with two posts each, one random post & Post3, Post4 -- Chained; Post5 -- Random \\ \hline 
  Incoherent & (2) & (1),(1),(1) & Intruded –- one subject, others random or different topics & Post3, Post4, Post5 - Intruded \\ \hline 
  Incoherent & None & (1),(1),(1),(1),(1) & All Random & Random \\ \hline
\end{longtable}

\textbf{Deciding which group caused the issue:}
If there are two groups of posts with the same size, please read through the available list of topical terms in order to determine which is the main topic of discussion, and note of the post numbers of the other group as the one causing the issue.

\subsection{Task Three: label the topics}
If a coherence score of 3 (Coherent) or 2 (Average) has been applied to the topic, please provide a high-level label (a word or a short phrase) for each topic, as follows:

\begin{enumerate}
    \item In the Topic label/s column, provide high-level labels (a word or a short phrase) for each topic. Please provide only one label for topics rated as coherent and up to two labels for topics rated as average.

\item Do not complete this task for topics with coherence scores of 1 (Incoherent). Where this applies, please note \textbf{(N/A)} in both columns. 
\end{enumerate}

\subsection{Task Four: Judge the relationship between the main topic and subtopics}

Once the previous three tasks have been completed for the main topic and a respective subtopic, the objective of this task is to judge the relatedness between the main topic and the subtopic. In the column ``Relatedness to the main topic``, assign a relatedness score according to the following scale:
\begin{enumerate}
    \item Assign a score of 3 (Strongly related) where the main topic and the subtopic are discussing related topics. For example, a point of discussion may have been mentioned in the main topic with more detail provided in the subtopic, or the subtopic has focused on an issue or topic with a strong connection to the main topic (regardless of whether it has been mentioned in the main topic). 
\item Assign a score of 2 (Partially related) where the main topic and the subtopic are somewhat related; for instance, if you believe the topics might be related but the relatedness is not readily apparent, direct, or comprehensive. 

\item Assign a score of 1 (Not related) where the main topic and the subtopic are not related and discuss different topics.

\item Assign a score of 0 (Identical) where the main and subtopic are identical and discussing exactly the same subject. For use when the subtopic does not provide any new information or nuance to the discussion of the topic under review, or where you find it difficult to distinguish between the main and the subtopic.
\end{enumerate}

\textbf{Dealing with average and incoherent topics}
In circumstances where a main topic or subtopic has been rated as incoherent, rate the relationship as (Not related). Furthermore, in instances where a topic has been rated as average with more than one label, and only one of the labels appears to be related, rate the relationship as (Partially related). 

\newpage
\section{Appendix B: Terms Extraction Results}
\label{AppendixB}

\begin{longtable}{
|p{\dimexpr0.7cm-2\tabcolsep-1.13\arrayrulewidth}
|p{\dimexpr3.3cm-2\tabcolsep-1.13\arrayrulewidth}
|p{\dimexpr4.2cm-2\tabcolsep-1.13\arrayrulewidth}
|p{\dimexpr4.2cm-2\tabcolsep-1.13\arrayrulewidth}
|p{\dimexpr4.2cm-2\tabcolsep-1.13\arrayrulewidth}|
}
\caption{Terms extraction results.}
\\\hline
  &
  \textbf{Topic label} &
  \textbf{Common terms in both 13- and 17-topic LDA} &
  \textbf{Terms unique to 13-topic LDA} &
  \textbf{Terms unique to 17-topic LDA }
   \\ \hline
\multicolumn{5}{|c|}{\textbf{Topics appeared in both 13- and 17-topic LDA.}} \\ \hline
\textbf{1} &
  Job requirements &
  experience, native,   degree, tefl, esl, course, company &
  certificate &
  country,   speaker, language, live, hire, require 
\\  \hline
\textbf{2} &
  Hiring process &
  interview, apply &
  referral, link, process, code &
  email, profile, application 
  \\ \hline
\textbf{3} &
  The class and the students &
  kid, student, level, lesson, class, time, call, teach &
  video, slide, read, conversation &
  child, late, show, start, camera, wait, young 
  \\ \hline
\textbf{4} &
  Bookings   and working hours &
  schedule, class,   book, slot, hour, time, week, day, month, open, weekend, booking
  &
  - &
  leave,   cancel, bonus, trial, regular, ph, cancelation
\\ \hline
\textbf{5} &
  Payments &
  rate, base, pay, low, make &
  hire, high, offer &
  price, tax, per 
  \\ \hline
\textbf{6} &
  Rating   system &
  rating,   give, feedback, review, bad &
  star &
  parent,   comment, assessment, good
\\  \hline
\textbf{7} &
  Technical issues &
  app, computer &
  issue, problem, try, test,   connection, 
  internet, email, send, post, check, & camera 
  \\ \hline
\multicolumn{5}{|c|}{\textbf{The new topic, absent from GT}} \\ \hline
\textbf{8} &
  Bank   transfers and transaction fees &
  bank, account, pay, paypal, payment &
  money, platform, price, charge &
  transfer, payoneer, fee  \\ \hline
\multicolumn{5}{|c|}{\textbf{Topics appeared only in either 13- or 17-topic LDA}} \\ \hline
\textbf{9} &
  The new contracts &
  N/A &
  contract, rating, new,   change, year, start   
  &
  N/A 
  \\  \hline 
\textbf{10} &
  How   tutors consider this job &
  N/A &
  work, live, job, time, money, need, life, income
  &
  N/A 
\\  \hline
\textbf{11} &
  Teaching material and methods &
  N/A &
  N/A &
  use, question,   conversation, learn, ask, slide, talk, answer, level, write, read 
  \\  \hline 
\textbf{12} &
  Reasons   to join or leave a platform &
  N/A &
  N/A &
  job, work, pay, make, money, online, business
\\  \hline
\textbf{13} &
  Miscommunication with platform management &
  N/A &
  N/A &
  contact, ticket, response,   email, send 
  \\ \hline
\multicolumn{5}{|c|}{\textbf{The missing topic from both LDA models}} \\ \hline
\textbf{ 14} & 
Covid-19-related discussions  & 
\multicolumn{3}{|c|}{Proposed terms:pandemic,Covid-19,lockdown}
\\ 
\hline
\end{longtable}

\newpage
\section{Appendix C: Phase Three (1) Codes and categories development} 
\label{AppendixC}  

\subsection{Focused codes and sub-codes after line-by-line coding}
{\scriptsize
\centering
\setlength\LTleft{-0.4in}
\begin{longtable}{
|p{\dimexpr1.6cm-1\tabcolsep-1.13\arrayrulewidth}
|p{\dimexpr1.7cm-1\tabcolsep-1.13\arrayrulewidth}
|p{\dimexpr6.5cm-1\tabcolsep-1.13\arrayrulewidth}
|p{\dimexpr6.5cm-1\tabcolsep-1.13\arrayrulewidth}|
}
\caption{Topic labels provided by the annotators, number of subtopics, and GT codes after line-by-line coding.}
\\\hline
\textbf{Topic Number} &\textbf{ Number of Sub-topics} & \textbf{Annotators labels }&   \textbf{Focused codes and sub-codes after line-by-line coding} \\ \hline
\endhead

1 & 5 & \begin{tabular}[c]{@{}l@{}}\textbf{Main}: Qualifications/Language qualifications /\\ Qualifications\\ \\ \textbf{Sub-topic1}: ESL companies/ESL\\ Companies/ Companies\\    \\ \textbf{Sub-topic2}:TEFL or CELTA/ \\ Teaching qualifications/ Training\\    \\ \textbf{Sub-topic3}:Countries/Preferred countries\\/ Countries\\    \\ \textbf{Sub-topic4:} Courses/  N/A   /Courses\\    \\ \textbf{Sub-topic5}: Native speakers/Native\\ language /Nationality \end{tabular}  & \begin{tabular}[c]{@{}l@{}}\textbf{Platforms polices and conditions}\\Deciding   factors to choose a platform\\Native speakerism and accent\\ Tours’ countries   and nationality \\Sacrificing  \\ \textbf{Redditting}:\\Explaining   the qualifications \\ Finding   the right company\\Empowering\end{tabular} \\ \hline

2 & 3 & \begin{tabular}[c]{@{}l@{}} \textbf{Main}: Recruitment process/ Recruitment / \\ Application process \\ \\\textbf{ Sub-topic1}: Application rejected/\\Application declined/Rejection\\ \\ \textbf{Sub-topic2}: Referral codes/Referrals/ N/A \\ \\ \textbf{Sub-topic3}: Interview process\\ /Interview demos/Interviews\end{tabular}  & \begin{tabular}[c]{@{}l@{}}\textbf{Platforms polices}\\ \textbf{Payments}\\\textbf{Confusion}\\\textbf{Redditting}:Creating a supportive network\\ Finding a ESL company \\ Seeking help in hiring process - Novices\\ Sharing   experiences - Experts\\ Report   being rejected\\ \textbf{Advertising referral codes} \\ \textbf{Strategising}\\ Applying using referrals\end{tabular} \\ 
\hline

3 & 2 & \begin{tabular}[c]{@{}l@{}}\textbf{Main}: Time management / Teaching \\ / Teaching levels\\    \\ \textbf{Sub-topic1}: Reservations/ PH calls/ \\ Overlapping\\    \\ \textbf{Sub-topic2}:N/A / Class timing/Scheduling\end{tabular} & \begin{tabular}[c]{@{}l@{}}\textbf{Technological systems }\\ Issues with classes scheduling\\ \& reservations:\\ Student showing up late/no-show,\\ Overlapping,\\ back-to-back reservations\\ Penalties\\ \textbf{Solving}:\\Time management\\Communication\\ \textbf{Continue tutoring} \\ Being patient\end{tabular} \\ \hline

4 & 5 & \begin{tabular}[c]{@{}l@{}}\textbf{Main}: Booking rates/Attracting bookings\\/Bookings\\    \\ \textbf{Sub-topic1}: Priority hours/Maximising \\  availability/Reservations\\    \\ \textbf{Sub-topic2}: Slots/Filling slots/Slots\\    \\ \textbf{Sub-topic3}: Weekends and holidays/\\Class cancellations/Holidays\\    \\ \textbf{Sub-topic4}: Bonuse / Contractual obligations \\/Pay     \\ \textbf{Sub-topic5}: Peak hour/Best timing/Hour\end{tabular} &  \begin{tabular}[c]{@{}l@{}}\textbf{Technological systems} \\Issues   with classes scheduling\\ \& reservations:\\ Overlapping\\Income instability\\Cancellations\\Changing   demand\\ \textbf{Confusion}\\\textbf{Redditting}: \\Seeking   help and explanations\\ Sharing experiences:\\ In managing schedule\\ In taking time off or cancelling classes \\ In contacting the platforms’management.\\ Sharing current   bookings status.\\ Sharing feelings\\Explaining the system\\\textbf{Continue tutoring}\\  Being patient\\  \textbf{Strategising}:\\ Working with   regular students\\ Maximizing   availability\\ Working   on multiple\\ platforms (multi-homing)\\ \textbf{Generating income}\end{tabular} \\ \hline

5 & 4 & \begin{tabular}[c]{@{}l@{}}\textbf{Main}: Pay/ Pay rates/ Bonuses\\    \\ \textbf{Sub-topic1}: Salary structure/Rates /\\Pay breakdown\\    \\ \textbf{Sub-topic2}: Low pay/Low pay\\/Low income\\    \\ \textbf{Sub-topic3}: Price of lesson/Class Price\\/Price\\    \\ \textbf{Sub-topic4}: Stability factor/Stability/ \\ Earnings adjustments\end{tabular} &  \begin{tabular}[c]{@{}l@{}}\textbf{Financial   uncertainty}\\ Max   earning, stability factor, etc.\\ \textbf{Unfairness:} Penalties\\\textbf{ Inadequate payment}\\ \textbf{Job Insecurity}- pay cuts\\ \textbf{Vagueness}\\\textbf{Solving}\\ \textbf{Strategising}\\multi-homing\\ platform   hopping: move to platforms \\allow tutors to set their own price.\\ \textbf{Redditting}: \\ Empowering\end{tabular} \\ 
\hline

6 & 1 & \begin{tabular}[c]{@{}l@{}}\textbf{Main}: Ratings/ Rating system/ Ratings\\    \\ \textbf{Sub-topic1}: Bad ratings/Managing \\ ratings/ Calculations \end{tabular} &  \begin{tabular}[c]{@{}l@{}}\textbf{Technological systems:} \\\textbf{Rating system issues} \\ Unfairness\\ Annoyance\\\textbf{Financial   uncertainty}\\ Penalties\\\textbf{Strategising}\\Protecting   reputation\\ \textbf{Confusion}\\ \textbf{Redditting}:\\ Seeking   advice \\Understanding and support:\\ Bonding over negative emotions\\ Explaining   the system \\sharing   experiences \\Empowering\end{tabular} \\ 
\hline

7 & 4 & \begin{tabular}[c]{@{}l@{}}\textbf{Main}: Internet/ IT problem/ Internet\\    \\ \textbf{Sub-topic1}: Technical problems\\/Technical issues /Connections\\    \\\textbf{ Sub-topic2}: Internet speed/Speed tests\\/speed\\    \\\textbf{ Sub-topic3}: Camera and audio/Mic and \\ headphone/Webcam\\    \\ \textbf{Sub-topic4}:App/New app/App\end{tabular} &  \begin{tabular}[c]{@{}l@{}}\textbf{Technical barriers}\\ \textbf{Financial   uncertainty}\\Penalties   \\  \textbf{Negative Feelings}\\\textbf{Solving}:\\ Contacting   platform management:\\\textbf{Strategising}\\ \textbf{Confusion}\\ \textbf{Redditting}:\\ Seeking explanations\\ Sharing   experiences \\Explaining   the system\end{tabular}  \\ 
\hline

8 & 3 & \begin{tabular}[c]{@{}l@{}}\textbf{Main}: Bank transfers/Financial fees/ Exchange\\    \\ \textbf{Sub-topic1}: Overseas transactions\\/Transfer /Payment platforms\\    \\ \textbf{Sub-topic2}: Transfer money/Currency/\\ Payment accounts \\    \\\textbf{ Sub-topic3}: Payment processes/Payment\\ methods/Payments\end{tabular} &  \begin{tabular}[c]{@{}l@{}}\textbf{Financial   uncertainty}\\Not   receiving the payment\\High costs\\\textbf{Redditting:} \\Seeking   help in bank transfers\\Explain   the transfer process and fees\\Sharing experiences.\\ \textbf{Vagueness}\end{tabular} \\
\hline

9 & 1 & \begin{tabular}[c]{@{}l@{}}\textbf{Main}: New contract/New contracts\\/Contracts    \\    \\ \textbf{Sub-topic1}: Contract terms/Contract \\changes / Conditions\end{tabular} &  \begin{tabular}[c]{@{}l@{}}\textbf{Financial uncertainty}\\ Pay cuts in new contracts\\ Tutors oversupply\\ Power-imbalance: Forcing to sign\\ Accounts termination \\ Sharing experience\\ Explaining the payment structure\\ \textbf{Confusion} \\ \textbf{Leaving} \end{tabular}  \\ 
\hline

10 & 2 & \begin{tabular}[c]{@{}l@{}}\textbf{Main}: Pay conditions/ Income / \\ Sustainable job \\ \\ \textbf{Sub-topic1}: Gig economy/Work \\ fluctuation /Hours\\    \\ \textbf{Sub-topic2}: Main income/Income \\source/Earning\end{tabular} &  \begin{tabular}[c]{@{}l@{}}\textbf{Being in need}\\The pandemic\\  Losing or changing   jobs.. etc. \\ Doing the job part-time or full time\\ Work-life balance\\Mental health \\ Bonding over negative emotions \\\textbf{Understanding}\\The business model of the gig economy  \\ \textbf{Managing expectation}\end{tabular}\\ 
\hline

11 & 7 & \begin{tabular}[c]{@{}l@{}}\textbf{Main}: Lessons/classes/Teaching approaches\\    \\ \textbf{Sub-topic1}: Conversation topics/ \\Conversation classes/ Topics\\    \\ \textbf{Sub-topic2}: Spoken language/Speaking/ \\ Learning strategies\\    \\ \textbf{Sub-topic3}: Problem student/Teaching\\ tips/Participation\\  \textbf{Sub-topic4}: Technical tools/Useful \\resources/Sharing\\    \\ \textbf{Sub-topic5}: Exam preparation/ Exams/\\ Helping students \\   \\ \textbf{Sub-topic6}: Learning a language/Learning/ \\Student experience\\  \textbf{Sub-topic7}:Teaching materials/Conversation\\ topics/Articles\end{tabular} &  \begin{tabular}[c]{@{}l@{}}\textbf{Redditting:}    \\ Seeking help in teaching\\ resources and methods\\ Sharing teaching resources\\ and methods\\Explaining platforms’ features \\and teaching rules\\Being a supportive and  caring colleague\\ Empowering \\\textbf{Confusion}\\ \textbf{Technological systems}\\ Reputation\\ \textbf{Strategising}\\ \textbf{Rising Popularity}\\Teaching popular lessons \\ Being likable  tutor\end{tabular}  \\ 
\hline

12 & 0 & \begin{tabular}[c]{@{}l@{}}\textbf{Main}: Business structure/Employee relation \\/ Companies \end{tabular} & \begin{tabular}[c]{@{}l@{}}\textbf{Understanding }\\The business model\\  Power imbalance\end{tabular} \\ 
\hline

13 & 2 & \begin{tabular}[c]{@{}l@{}}\textbf{Main}: Technical problem/ Issue/\\ Company communications    \\ \textbf{Sub-topic1}: Email problems/Missing\\ emails/Emails\\    \\ \textbf{Sub-topic2}: Absences/Absent marks\\/Absences\end{tabular} &  \begin{tabular}[c]{@{}l@{}}\textbf{Technological systems:} \\ Technical barriers\\Penalties \\ \textbf{Miscommunications} with the management\\\textbf{Solving}:\\  Communications \\ appealing penalties\\ \textbf{Confusion}\\\textbf{Redditting}: \\Explaining the absence system\end{tabular} \\ \hline

14 & 2 & \begin{tabular}[c]{@{}l@{}}\textbf{Main}: Online teaching/Teaching \\opportunities/ Pandemic \\    \\ \textbf{Sub-topic1}: Employment prospects\\/Employment strategies/ Careers\\    \\ \textbf{Sub-topic2}: Marketing/Marketing / \\Teaching ventures \end{tabular} &  \begin{tabular}[c]{@{}l@{}}\textbf{Understanding}:\\The business model\\Pandemic effect \\ \textbf{Strategising}\\\textbf{Solving}:\\ Investing in professional development\\Self-Marketing\\ \textbf{Leaving} then freelancing as an online tutor\end{tabular} \\ \hline 

\end{longtable}
}

\newpage
\subsection{The Core categories}
\begin{table}[h]
\caption{Categories created after comparing the focused codes and sub-codes of the 14 groups of topics.}
\centering
{\scriptsize
\begin{tabular}{|l|l|l|}
\\\hline
 & \textbf{Category}                      &\textbf{ Focused codes and sub-codes} \\\hline                                                                                     
1 \multirow{3}{*}{} &
  \multirow{3}{*}{\textbf{Under-structured organisations}} &
  \begin{tabular}[c]{@{}l@{}}Polices\\ Power \\ Lack of communication \\ \\ \end{tabular} \\
  
 &                               & \begin{tabular}[c]{@{}l@{}}\textbf{Financial uncertainty}:\\ Income instability\\ Job insecurity\\ Inadequate payments \\ \\  \end{tabular}  \\
 
 &                               & \begin{tabular}[c]{@{}l@{}}\textbf{Technological systems}:\\ Technical issues\\ Booking system\\ Rating system\end{tabular}         \\ \hline
2  & \textbf{Confusion}                     &                           \\\hline                                                                                                 
3 & \textbf{Being in need }                & \begin{tabular}[c]{@{}l@{}}Losing job\\ Studying toward a degree\\ Continue tutoring\end{tabular}                          \\ \hline
4 & \textbf{Financial structure}           & Doing the job full-time or part-time                                                                                       \\ \hline
5 &\textbf{Competency-level}             & Novices and Experts                                                                                                        \\ \hline
6 &
  \multirow{3}{*}{\textbf{Persisting}} &
  \begin{tabular}[c]{@{}l@{}}\textbf{Redditting}: \\Creating supportive network\\ \textbf{Competency stages’ individuals roles:} \\ Help seekers - Novices\\ Help providers - Experts:\\ Offering explanations and solutions\\ Sharing experiences and current status\\
  
  \\ \textbf{Resisting}: \\ Bonding over negative emotions\\ Empowering \\ Being patient \\  \\  \end{tabular} \\
 &                               & \begin{tabular}[c]{@{}l@{}}\textbf{Solving}: \\ Contacting platforms management\\ Appealing and providing proofs \\ Education students about the systems\\Investing in professional development \\Time management\\ Self-marketing \\ \\ \end{tabular}                        \\ 
 &
   &
  \begin{tabular}[c]{@{}l@{}}\textbf{Strategising}:\\ Building base of regulars \\ Maximizing availability \\ Protecting reputation \\ Rising popularity\\ Multi-homing \\ Applying using referral codes\end{tabular} \\ \hline
 7 & \multirow{4}{*}{\textbf{Consequences}} &\begin{tabular}[c]{@{}l@{}}\ \textbf{Generating income} \\ \textbf{Realising} \\ \end{tabular}
                                                             \\
 &                               & \begin{tabular}[c]{@{}l@{}}\textbf{\textbf{Staying:}} \\ Managing expectations \\Sacrificing \end{tabular}                                                  \\
 &                               & \begin{tabular}[c]{@{}l@{}}\textbf{\textbf{Leaving:}}\\ Platform-hopping\\ Freelance as online tutors\end{tabular}                     \\ \hline
\end{tabular}%
}

\label{Categories_table}
\end{table}

\newpage
\section{Appendix D: Phase Three (2) Samples of empirical data and corresponding codes }

\label{AppendixD}  

These tables offer examples of incidents from the data related to each of the core categories in our theory and their corresponding codes.

{\scriptsize

\begin{longtable}{|p{0.13\textwidth}|p{0.85\textwidth}|}
\caption{The condition: Working with under-structured organisations}
\\\hline

\textbf{Code} & \textbf{Sample of incident from the data} \\
\hline
\endfirsthead

\hline
\textbf{Code} & \textbf{Sample of incident from the data} \\
\hline
\endhead

\textbf{Job Insecurity} &
\begin{tabular}[t]{@{}p{0.85\textwidth}@{}} 
``Part of being a freelancer is always worrying about job security. Definitely not the best part of the job''\\ ``I know no job is perfect, and I'm mostly concerned over job security''\\ ``The thought of getting fired at any given moment, thus becoming homeless, is really stressing me out. I'm praying I can hold on until September when I go back teaching for real and for decent money.''\\ ``My accounts have suddenly been deactivated without any reason after a few months teaching on Cambly.''\\ ``They reassured us that everything would remain the same   {[}...{]}. Now, 3 months later, they are forcing us into contracts with lower pay under the threat of immediate termination. So, yes, it is bullying''\\ ``I just received an email stating that they are giving me a final chance to sign the new contract, and if I don’t, they’ll terminate my account permanently''\\ ``Because there are over 30,000 of us. We're easily   replaceable''\\ ``I'd say the market is super saturated and you will get very low pay''\end{tabular} \\ \hline

\textbf{Income Instability} &
\begin{tabular}[t]{@{}p{0.85\textwidth}@{}}

``My earnings this financial year are 40\% of last year due to the saturated English teaching market since COVID''\\ `` I’m worried that the online market is over-saturated since so many people are currently moving to online teaching''\\  ``It seems like every single Online ESL Company is oversaturated at the moment, I have a five-star rating and still have trouble obtaining a full schedule''\\  ``Children are back at school, this mean low demand ''\\ ``Students are dropping like flies, they always come and go, it is hard not to feel a bit of instability'' \\ ``Online ESL is feast and famine. People that get on board in the feast times, usually summer and November \& December''\\ ``The instability is just too much now'' \\ ``This morning one of my reservations was 8 minutes late and when he finally called, he had a poor connection, abruptly hung up after 2 minutes, and never called back   {[}..{]} I then only get paid for the 2 minutes''\\ `` I had to   sit in front of my computer for a few days with no classes''\\ `` we're still expected to sit in front of a computer not being paid waiting for sudden   classes that never come''\end{tabular} \\ \hline

\textbf{Inadequate monetary compensation} &
\begin{tabular}[t]{@{}p{0.85\textwidth}@{}}
``The base pay is very low, and I'm not happy with that''\\  ``In most places, you could earn more by claiming unemployment'' \\ ``It's just not worth the mental energy for such low pay''\\ ``The pay is   already horrifically low, and I pay 33\% of the already low pay in taxes'' \\ ``Sending money in USD and then transferring it to Europe results in significant fees for both the currency exchange and  the transfer itself''\end{tabular} \\ \hline

\textbf{Bookings system} &
\begin{tabular}[t]{@{}p{0.85\textwidth}@{}}
``She booked two time slots back to back just to cancel all of them. It blocks the time slots for other students and messes with my day''\\ ``Tomorrow I've got seven reservations and I just know I'll get at least two no-shows or last-minute cancellations. It always happens''\\ ``We have to be online but aren't guaranteed any minimum pay .. on no-shows as you will be paid for 10 minutes of a confirmed lesson''\\ ``Yesterday, a student cancelled 45 minutes before the due lesson, and I received a paltry   \$1.87, saving them \$5.63. I should be entitled to at least half that!''\\ ``Ok, so I need some help because now I'm getting a little worried about the reservation/priority hour overlap. I don't want to be penalised because my reservation and PH (priority hour) is coming up soon. I don't know how to handle the overlap. Freaking out just a little   here, Please help!'' \\ ``Sometimes, a  couple of your students experience an internet connection issue during the class, and the reconnecting process adds an additional 60 seconds for those two classes. All these things are normal circumstances leading the 5th and 6th classes to be 5 minutes late. However, students definitely won't be happy   about paying for a 30-minute class and getting only a 28-minute class due to   no fault of their own''\end{tabular} \\ \hline

\textbf{Rating system} &
\begin{tabular}[t]{@{}p{0.85\textwidth}@{}}
``Your ratings by students determine whether or not you receive the Performance Bonus'' \\ ``It was more likely due to your low rating. I think they probably have an automated system that gets rid of any tutors below a certain rating''\\  ``Many students started calling and telling me 'I called you because you have a 5* rating'''\end{tabular} \\ \hline

\textbf{Unfairness} &
\begin{tabular}[t]{@{}p{0.85\textwidth}@{}}

``the fact that your overall rating, and therefore booking capacity and livelihood, is at the mercy of a 5 year old figuring out a rating system is completely absurd and unfair'' \\ ``I agree that the rating system is completely unfair, and they should either end it or, at the very least, make sure students give us some kind of feedback so we know how to improve'' \\ ``You will get bad ratings for not understanding the material, bad connection, noisy background, or if they had the best tutor before and then they had your class, etc''\\ ``I actually had something similar happen. It might be a bug, and I got a low rating because of it''\\ ``I've stopped caring about my ratings because you can work very hard for a high rating, and just a single low rating can ruin everything''\\ ``Their rating system drives me nuts. Mine is decent (4.87), down from 4.90 recently. I'm tired of my income depending on an arbitrary rating system.''\\  ``I was   deactivated!, I think they probably have an automated system that gets rid of any tutors below a certain rating. They can get rid of us whenever they want''\end{tabular} \\ \hline

\textbf{Technical barriers} &
\begin{tabular}[t]{@{}p{0.85\textwidth}@{}}
``I'm losing my passes because of issues on their end, and I'll eventually run out and have to pay penalties \\ or even get fired. It feels like they're not taking me seriously'' \\ `` It is so annoying to go through such a   process to be believed, but I'm in that habit it now ton-sometimes-get an appeal to go through'' \\ ``I've been kicked out for no reason and couldn't get back in, I tried appealing the absent for 6 times and even sent screenshots. They would not invalidate it. It is certainly infuriating'' \\ `` I've submitted numerous appeals along with screenshots and evidence of having a stable connection and have been talking in circles with the comm center and they refuse to remove \\ any of the absences, or acknowledge the problem is on their end''\end{tabular} \\ \hline

\textbf{Failing to comply to policies} &
\begin{tabular}[t]{@{}p{0.85\textwidth}@{}}
`` Right now we have a huge influx of new students coming to the company, so I don't think it's gonna be so hard for new people coming in. There's definitely a need for teachers'' \\ ``All these companies advertise more than they actually pay to get people in''\\ ``it's not what they advertise on their website''\\ ``Anyone have any suggestions for a decent company?  Even if the pay is a little   less than other companies, \\ I wouldn't mind as long as they treat you reasonably, and offer what they advertise''\end{tabular} \\ \hline

\textbf{Classification of workers} &
\begin{tabular}[t]{@{}p{0.85\textwidth}@{}}
`` There's no contract, no minimum weekly requirements, no strict rules, and you don't have to stick to \\ a schedule. You can just logon anytime you want to tutor''\\ `` You are required to teach minimum 8 fixed hours during Monday to Sunday 7PM-9PM per week''\end{tabular} \\ \hline

\textbf{Lacking of effective communication channels} &
\begin{tabular}[t]{@{}p{0.85\textwidth}@{}}
``There is poor communication with administration, and they do not respond efficiently'' \\ ``Their responses are a mix of robotic/totalitarian'' \\ ``Well.... I never heard back, but the strangest part is I never even got a 'no thank you' or anything, just zero response'' \\ ``I've emailed them four times and just keep getting the same reply over and over''\\ ``You can only send one message at a time''\end{tabular} \\ \hline

\textbf{Lacking of respect} &
\begin{tabular}[t]{@{}p{0.85\textwidth}@{}}
``There is a lack respect for teachers''\\ ``The new contract reduced   wages by 30\% - they provide no notice of holiday off time - they manipulate the system - especially the ratings to then send a standard email of punishment … just a shame they’re unable to provide any teacher respect'' \\ ``The fact that they didn't bother to inform tutors pretty well illustrates the level of respect they have for us''
\end{tabular}\\\hline
\end{longtable}
}

{\scriptsize
\begin{longtable}{|p{0.13\textwidth}|p{0.85\textwidth}|}
\caption{The cause: Being in need, and the context: Confusion}
\\\hline



\textbf{Being in need} &
\begin{tabular}[t]{@{}p{0.85\textwidth}@{}} 
``I lost my job due to COVID, unfortunately. Kind of have my back against the wall at the moment'' \\ ``Teaching online seems to be a   good way to keep myself afloat while I await global normality'' \\ ``I'm a full-time student and working on Cambly as my main source of income while I complete my degree''\\ ``I used to substitute and teach in a brick and mortar school, but now, since I am a stay-at-home mom, I   need to teach online to make some money'' \\  ``The money I am making is what I eat with and pay bills with``\\ ``I am doing it just as a side job to help save'' \\ ``it as a side job, a 2nd   income stream''\\  ``The thought of getting fired at any given moment, thus becoming homeless, is really stressing me out'' \\ ``How far out are you booked? I   hustled like crazy over the summer to keep my family afloat during a furlough   (thanks Covid), and then worked mostly weekends \& school breaks until   April'' \\ ``the inconsistency in bookings is stressful right now because it's my main income source'' \\  ``It’s hard not to worry about it because   it’s my main source of income''\end{tabular} \\ \\ \hline

\textbf{Confusion} &

\begin{tabular}[t]{@{}p{0.85\textwidth}@{}} 
``We get no training''\\ ``I'm not really understanding how this work''\\ ``Now, I have one absence in my record. What should I do? Should I use the free pass? Will using the free pass erase my absence from my record? Please help!'' \\ ``Do we only get paid if they actually show up? Confused as to how this works, exactly''\\ ``I'm very confused by this and quite disappointed really as I thought I was doing a good job''\end{tabular} \\ \\ \hline
\end{longtable}}

{\scriptsize
\begin{longtable}{|p{0.13\textwidth}|p{0.85\textwidth}|}
\caption{Co-variances: (Competency level and Financial structures) and the contingency: Continue tutoring}
\\\hline

\textbf{Code} & \textbf{Sample of incident from the data} \\
\hline
\endfirsthead

\multicolumn{2}{c}{{\bfseries Table \thetable\ continued from previous page}} \\
\hline
\textbf{Code} & \textbf{Sample of incident from the data} \\
\hline
\endhead

\textbf{Novice:}New to platform, expert in teaching &
\begin{tabular}[t]{@{}p{0.85\textwidth}@{}} 
``I've been thinking about doing some online teaching as a way to pick up some extra money when I'm not busy in the mornings. I've got over 10 years of teaching experience with lots of experience doing one-to-one instruction'' \end{tabular} \\ \hline
\textbf{Novice:} New to platform, new to teaching &
\begin{tabular}[t]{@{}p{0.85\textwidth}@{}} 
  ``I am currently setting up an account on Cambly. I'm completely new to teaching so appreciate these insights.''\end{tabular} \\ \hline
\textbf{Novice:} Unfavorable technological systems &
\begin{tabular}[t]{@{}p{0.85\textwidth}@{}} 
`` I am a newbie and realised that you just need one bad rating for your overall average to drop, and then it is very difficult to get it back up''\\ ``my average rating isn't very high and it was mainly because I just started and hadn't had a lot of classes yet and got one bad rating :( I still haven't been able to increase my average to higher than 9. It's sort of stagnating at 8.77 right now''\end{tabular} \\ \\ \hline
\textbf{Expert:} Time - what to expect &
\begin{tabular}[t]{@{}p{0.85\textwidth}@{}} 
``I've been working with VIPKID for 3 years now {[}…{]} If you're teaching in this time zone with full bookings, you'll likely have 10-11 classes a day.''\\ ``I've been with Dada now 1257 days according to the app.''\end{tabular} \\ \hline
\textbf{Expert:} Established base of regulars &
\begin{tabular}[t]{@{}p{0.85\textwidth}@{}}
  ``It’s my first contract. I have taught nearly 1,000 classes over the five months I’ve been here. Currently have 44 students(most of whom do regularly book) and several more recurring.'' \end{tabular} \\ \\ \hline
\textbf{Expert:} High rating and pay &
\begin{tabular}[t]{@{}p{0.85\textwidth}@{}} 
``I've worked with them for close to a year now, my rating is good and so is my pay'' \end{tabular} \\ \hline
\textbf{Financial structures} &
\begin{tabular}[t]{@{}p{0.85\textwidth}@{}} 
``I make a decent living, had support from my family and when we lived together, contributed mutually to the few bills we pay. I really enjoy teaching so for me, it works''\\ ``Yes, I do this job full-time, but I have financial support''\\ ``I live in a low cost of living country so the pay is pretty good for here'' \\ ``I feel like my whole day revolves around this job, which is depressing''\\ ``If my main job can cover my mortgage, food, etc., I would give up an additional job that made me stressed and severely depressed''\end{tabular} \\ \\ \hline
\textbf{The Contingency: Continue tutoring} &
\begin{tabular}[t]{@{}p{0.85\textwidth}@{}} 
``I, for one, cannot leave as I need this income, despite how bad I find the job''\\ ``My pay has dropped due to the INCREDIBLE amount of cancels. I have also lost a few students, but I expected that due to COVID and the global economy taking a hit. Things are just weird from all angles, but I'm going to hang in until the bitter, bitter end''\end{tabular} \\ \\ \hline

\end{longtable}}

{\scriptsize
\begin{longtable}{|p{0.13\textwidth}|p{0.85\textwidth}|}
\caption{The behaviour: Persisting}
\\\hline
\cline{1-2}
\multicolumn{1}{|l|}{\textbf{Code}} & \textbf{Sample of incident from the data} \\ \cline{1-2}
\endfirsthead
\multicolumn{2}{c}{{\bfseries Table \thetable\ continued from previous page}} \\
\cline{1-2}
\multicolumn{1}{|l|}{\textbf{Code}} & \textbf{Sample of incident from the data} \\ \cline{1-2}
\endhead

\multicolumn{2}{|l|}{\textbf{Redditing}} \\ \cline{1-2}
\multicolumn{1}{|p{0.13\textwidth}|}{\textbf{Sharing feelings}} &
\begin{tabular}[c]{@{}p{0.85\textwidth}@{}}
``I'm excited to start teaching''\\  ``The bad ratings still annoy me when they happen.'' \end{tabular} \\ \cline{1-2}
\multicolumn{1}{|p{0.13\textwidth}|}{\textbf{Help seeking}} &
\begin{tabular}[c]{@{}p{0.85\textwidth}@{}}
  ``I just need some clarification with how the schedules work.''\\  ``I just wanted to ask a question about teaching strategies and the best way to teach low-level learners''  \end{tabular} \\ \cline{1-2}
\multicolumn{1}{|p{0.13\textwidth}|}{\textbf{Explaining the system}} &
``You can open your reservations to anyone or (as I do) you can set your reservations to  "regulars only". ''  \\ \cline{1-2}
\multicolumn{1}{|p{0.13\textwidth}|}{\textbf{Offering solutions}} &
  ``Try to use other equipment eg. headset, may solve the issue.'' \\ \cline{1-2}
\multicolumn{1}{|p{0.13\textwidth}|}{\textbf{Sharing experience}} &
  ``**from my experience**. I took 3 days off, which for me is  about 14 hours off, made the request before classes scheduled with no penalty''\\ \cline{1-2}
\multicolumn{1}{|p{0.13\textwidth}|}{\textbf{Sharing current financial status}} &
  ``I make around \$1400 per month and I started in January. I only work about 20 hours a week'' \\ \cline{1-2}
\multicolumn{1}{|p{0.13\textwidth}|}{\textbf{Sharing schedule status}} &
  ``I teaching adults only, I do about 30 half hour sessions per week. So for 15 hours a week, that's not bad. I've been with Cambly for about 6 months now.'' \\ \cline{1-2}
\multicolumn{2}{|l|}{\textbf{Resisting}}  \\ \cline{1-2}
\multicolumn{1}{|p{0.13\textwidth}|}{\textbf{Bonding over negative emotions}} &
\begin{tabular}[c]{@{}l@{}}
``I understand how stressful it can be to have a tech issue during demos. \\This threw me when I was new as well!''\\
`` Whatever your feeling is totally justified. I’ve been working for 2 years for this company \\ and this is the most unstable it’s ever has been.''\\  ``I think there isn’t much you can do in these situations but I know how frustrating it can be''\end{tabular}  \\ \cline{1-2}
  
\multicolumn{1}{|p{0.13\textwidth}|}{\textbf{Empowering}} &
\begin{tabular}[c]{@{}p{0.85\textwidth}@{}}
``When choosing a company to teach for it's not just about us meeting their needs, it's also about them meeting our needs by providing the pay, flexibility, policies, etc. that meet our needs.''\\   ``We're Independent Contractors that are Self-Employed, and can pick and choose which Online ESL companies we want to offer our services to. And there are a lot of Online ESL companies to choose from.'' \end{tabular}   \\ \cline{1-2}

\multicolumn{1}{|p{0.13\textwidth}|}{\textbf{Being patient}} &
\begin{tabular}[c]{@{}p{0.85\textwidth}@{}}
``General rule is that the more ratings you have the better a reflection it is of you. So yeah, hang in there''\\  ``I've gotten better over time. My score has gone up and bad ratings has gone down''\\ ``It takes time to get established on the platform and there are a lot of new tutors recently but if you stick it out you can do pretty well after a couple months of effort.''\\ ``If you're serious, consistent and passionate, you'll do fine. From what I've observed, people can make good money in this industry. But it takes time.'' \end{tabular} \\ \cline{1-2}
  
\multicolumn{2}{|l|}{\textbf{Solving}}  \\ \cline{1-2}
\multicolumn{1}{|p{0.13\textwidth}|}{\textbf{Contacting platforms management}} &
\begin{tabular}[c]{@{}p{0.85\textwidth}@{}}
``I wrote to comm center about this.''\\  ``I've sent an email to italki support''\end{tabular}  \\ \cline{1-2}
\multicolumn{1}{|p{0.13\textwidth}|}{\textbf{Appealing}} &
  ``Send them a message through the comm center or appeal page and let them know that your class disappeared.'' \\ \cline{1-2}
\multicolumn{1}{|p{0.13\textwidth}|}{\textbf{Providing proofs}} &
\begin{tabular}[c]{@{}p{0.85\textwidth}@{}}
``You can appeal your absence and upload them as evidence. Whenever something goes weird for me, I always screenshot just in case.''\\ ``Typically what you need is a screenshot of your internet speed test at the time of the incident, screenshots of you talking to tech support, and a screenshot of whatever went wrong.''\end{tabular}  \\ \cline{1-2}

\multicolumn{1}{|p{0.13\textwidth}|}{\textbf{Educating students about the system}} &
``I tell them the rating should only be about the teacher. Student should know what ratings mean to us, I think students are rating the class as a whole, which reflects bad on us given the horrible material we have to use and the issues we have with loud background noises.'' \\ \cline{1-2}

\multicolumn{1}{|p{0.13\textwidth}|}{\textbf{Investing in professional development}} &
  ``I want to get a certificate that makes it look like I know something, but people here said it is unlikely to learn much there.''  \\ \cline{1-2}

\multicolumn{1}{|p{0.13\textwidth}|}{\textbf{Time management}} &
\begin{tabular}[c]{@{}p{0.85\textwidth}@{}}
``I start each class on time of the student calls me on time. I politely tell the student who was late that I have to go because it is time for my next class to start.'' \\  ``In case of a client arrives really late I cover the vocab and skip some slides then try my best to give them an overview''\end{tabular}  \\ \cline{1-2}

\multicolumn{1}{|p{0.13\textwidth}|}{\textbf{Self-Marketing}} &
\begin{tabular}[c]{@{}p{0.85\textwidth}@{}}
``I was going to suggest you redo your intro video and update your bio. It might attract new parents/students. I got 2 students simply because I have cats.''\\  ``I get a lot of regular students from people who were browsing my profile and found something they connected with (usually from my "interests" section but sometimes with my education section or profession). If you can sell your   skills enough, students will go straight to reserve a time with you.''\end{tabular}  \\ \cline{1-2}

\multicolumn{2}{|l|}{\textbf{Strategising}}  \\ \cline{1-2}
\multicolumn{1}{|p{0.13\textwidth}|}{\textbf{Building base of regulars}} &
\begin{tabular}[c]{@{}p{0.85\textwidth}@{}}
``I have quite a few regular students they always schedule at the same time every week.``\\  ``Built up a base or regulars, I like it better than PH and the income is more steady as well. ''\end{tabular}  \\ \cline{1-2}
\multicolumn{1}{|p{0.13\textwidth}|}{\textbf{Maximizing availability}} &
``At first, you   need to be online as much as possible, working during peak hrs and open up all the slots for pre-booked classes.'' \\ \cline{1-2}
  
\multicolumn{1}{|p{0.13\textwidth}|}{\textbf{Rising Popularity: Teaching Popular type of   classes}} &
\begin{tabular}[c]{@{}p{0.85\textwidth}@{}}
``Try to get into the OUP. These classes are popular, during the week I have 4 OUP regular students which is nice. \\  Students on the platform specifically want IELTS tutors, If it's a bad day I may have to teach IELTS using the Cambly material! (because the student has requested it!).''\end{tabular}  \\ \cline{1-2}

\multicolumn{1}{|p{0.13\textwidth}|}{\textbf{Rising Popularity: Being a likable tutor}} &
\begin{tabular}[c]{@{}p{0.85\textwidth}@{}}
``My approach is to try to get students like me. This way you're guaranteed a good rating, just be a likable person.''\\ ``I'm not bubbly, but I joke around with the kids and smile. This always make them come back''\end{tabular} \\ \cline{1-2}

\multicolumn{1}{|p{0.13\textwidth}|}{\textbf{Protecting reputation: Loyalty}} &
\begin{tabular}[c]{@{}p{0.85\textwidth}@{}} 
``Work mainly with regular students, they repeatedly give you a good rating, your rating goes up, more students book reservations, building your regular base of students.''\\ ``Regulars always leaving 5 stars after every class, this is the one that really works to bring your score up.''\\ ``Set your reservations to "regulars only" instead. Usually your regulars will only leave you 5 star ratings.   Keep it that way until you are satisfied with your rating.''\end{tabular}  \\ \cline{1-2}

\multicolumn{1}{|p{0.13\textwidth}|}{\textbf{Protecting reputation: Being direct}} &
\begin{tabular}[c]{@{}l@{}}
``ALWAYS ask for a 5 star rating after EVERY class. I  have a nice, polite little script where \\ I humbly beg for stars. Have   no shame. Don't hesitate. (never just ask for stars, always 5 stars).''\\ ``I am asking   all PH students to give me 5 stars and I ask them on the call.''\end{tabular}  \\ \cline{1-2}
  
\multicolumn{1}{|p{0.13\textwidth}|}{\textbf{Multi-homing}} &
\begin{tabular}[c]{@{}p{0.85\textwidth}@{}}
``I work for two companies since bookings are so low and I have opened all slots in both. When a spot fills up with one, I close it for the other. If your schedule is full or if bookings fill up then you can adjust your method to reflect.''\\ ``I have 4 online teaching jobs and this keeps me afloat. Before the pandemic only had 1 job. ''\end{tabular} \\ \cline{1-2}

\multicolumn{1}{|p{0.13\textwidth}|}{\textbf{Applying using referrals}} &
\begin{tabular}[c]{@{}p{0.85\textwidth}@{}}
``It is better to apply using referral links, however, if you choose to apply directly through Qkids website instead of using the assistance of a Teacher/Recruiter, then you have to figure everything out on your own.''\\ ``Applicants who apply using a referral code have a better chance of getting hired.''\end{tabular}  \\ \cline{1-2}
\end{longtable}}

{\scriptsize
\begin{longtable}[c]{|p{0.2\textwidth}|p{0.8\textwidth}|}
\caption{The consequences}
\\\hline
\textbf{Code} & \textbf{Sample of incidents from the data} \\ \hline
\endfirsthead
\multicolumn{2}{c}%
{{\bfseries Table \thetable\ continued from previous page}} \\
\hline
\textbf{Code} & \textbf{Sample of incidents from the data} \\ \hline
\endhead
\textbf{Generating sufficient income} &
  ``I maintained a very good salary. Now that I can't keep up with the hours, I've eliminated all of the lower level kids and only kept the kids that I love or that fill a time and don't annoy me much.'' \\ \hline
\textbf{Realising} &
  ``I get why and it totally makes sense! That explains all my cancellations for today!'' \\ \hline
\textbf{Accepting} &
  ``It’s just business, that's how a business works.'' \\ \hline
\textbf{Managing expectations} &
  ``My advice would be is to manage your expectations. Don't think of it as something you can support a family with, but just a side gig with a bit of income.'' \\ \hline
\textbf{Sacrificing} &
  ``Indeed, with Cambly we sacrifice high pay for high flexibility. There are many other online teaching programs that pay more but they require consistent scheduling and mandatory weekend.'' \\ \hline
\textbf{Sacrificing: work-life balance} &
  ``If you're okay working a lot of weird hours and you aren't a high spender, it can work.'' \\ \hline
\textbf{Leaving} &
  \begin{tabular}[c]{@{}p{0.8\textwidth}@{}}
  ``I am trying to find a new job because tutoring online will no longer be sustainable as a full time gig.'' \\
  ``My bookings are horrible I’m thinking about leaving.''
  \end{tabular} \\ \hline
\textbf{Platform hopping} &
  ``I have recently accepted a new position at another company that pays more than what I made at my peak at Dada and I've been closing off my slots at Dada.'' \\ \hline
\textbf{Leaving then freelancing as an independent online tutor} &
  ``Just find your own clients and teach as an independent teacher. You can find good quality adult clients on LinkedIn if you've got a good online presence.'' \\ \hline
\textbf{Moving to platforms allow tutors to set their own price} &
  ``Try platforms where you advertise yourself and set your own price. I feel most students would pay more for a teacher they liked.'' \\ \hline
\textbf{Lowering their rates initially} &
  ``Just a point, when starting out, keep your prices low until you generate students and then you can raise your price later. If you put the scale down to the cheapest 4 teachers, I was one of them.'' \\ \hline
\textbf{Selling lesson packages} &
  ``Packages are useful for attracting long term, It's nice to have some consistency in your schedule.'' \\ \hline

\end{longtable}}

\end{sm}

\end{document}